\documentclass[a4paper,11pt]{article}
\usepackage{jcappub}

\graphicspath{{figures/}}

\usepackage{lipsum}
\usepackage{physics}
\usepackage{mathrsfs}
\usepackage{multirow}
\usepackage{xspace}
\usepackage{natbib}
\usepackage{fontawesome5}
\usepackage{xcolor}
\usepackage{wrapfig}
\usepackage{amsmath, amssymb}
\usepackage[figuresright]{rotating}
\usepackage[fontsize=11pt]{fontsize}
\usepackage{comment}
\usepackage{booktabs}
\usepackage{orcidlink}

\usepackage[hang,flushmargin]{footmisc}

\newcommand{\PIP}[1]{\textcolor{magenta}{#1}}
\newcommand{\matt}[1]{\textcolor{green}{#1}}

\newcommand{\hii}{\ensuremath{\mathrm{H\textsc{ii}}}}
\newcommand{\cii}{\ensuremath{\mathrm{C\textsc{ii}}}}

\newcommand{\nhat}{\boldsymbol{\hat n}}
\newcommand{\bfx}{\boldsymbol{x}}
\newcommand{\bfk}{\boldsymbol{k}}

\defcitealias{Petrovic_2011}{PO11}

\usepackage[
    starfontserif 
    ]{starfont}
\usepackage{amsmath}

\DeclareSymbolFont{starfontsym}{OT1}{sts}{m}{n}
\DeclareMathSymbol{\mathSun}{\mathord}{starfontsym}{115}

\makeatletter

\newcommand{\checknextarg}{\@ifnextchar\bgroup{\gobblenextarg}{}}
\newcommand{\gobblenextarg}[1]{\,\mathrm{#1}\@ifnextchar\bgroup{\gobblenextarg}{}}
\makeatother

\begin{document}

\title{Radio Recombination Line Contamination in Post-Reionization 21~cm Intensity Mapping}


\author[a]{Pip Petersen\,\orcidlink{0000-0002-6307-7932}}
\affiliation[a]{Department of Astronomy, University of Washington, 3910 15th Avenue NE, Seattle, WA 98195, USA}
\affiliation[b]{School of Physics and Astronomy, Tel Aviv University, Ramat Aviv 69978, Israel}

\author[a,b]{{Yakov Faerman}\,\orcidlink{0000-0003-3520-6503}}

\author[a]{{Matthew McQuinn}\,\orcidlink{0000-0001-7961-9735}}

\emailAdd{speter7@uw.edu}

\abstract{
    We explore radio recombination line (RRL) contamination in post-reionization 21~cm intensity mapping observations. We develop a formalism to estimate the contamination of the 21~cm auto-power spectrum from the myriad of RRL lines that redshift into a 21~cm map and predict contamination of the 21~cm power spectrum at all redshifts. At $z\lesssim 2$, extrapolation of the conservative upper-bound model of \citet{Petrovic_2011} predicts high contamination to the 21~cm power spectrum, whereas our empirically-calibrated models suggest contamination several orders of magnitude smaller. We also estimate the contribution from carbon RRLs and find it to be subdominant. We find the RRL contamination is highly oscillatory in wavenumber, owing to the dominant contamination from RRLs emitted close in physical space to the 21~cm emission. These oscillations could in principle bias distance measurements derived from Baryon Acoustic Oscillations (BAO), a key science driver of post-reionization 21~cm intensity mapping. We find the shift in the BAO extrema is at least two orders of magnitude smaller than relevant for percent-precision cosmology from BAO.} 


\maketitle
\flushbottom
\section{Introduction}
\noindent 

Measuring cosmological parameters to high precision depends on reliable mapping of as much of the observable universe as possible. The standard approach has been to conduct large-scale galaxy surveys, resolving the positions and distances of the brightest sources to the highest accessible redshifts. In recent years, intensity mapping has emerged as a promising strategy for measuring the large-scale structure of the universe at these redshifts, mapping the cumulative intensity of dim sources that would be too faint to detect with traditional galaxy surveys. In the last decade, many intensity mapping experiments have emerged, focusing on mapping the emission of many spectral lines (such as CO, $\cii$, H$\alpha$, and Ly$\alpha$) over a wide range of redshifts \citep{Kovetz_2017}. Among these experiments, 21~cm intensity mapping stands as the ideal method for measuring the evolution of neutral hydrogen through the Dark Ages, Epoch of Reionization (EoR), and post-reionization for precision cosmology \citep{Furlanetto_2006, Morales_2010}. This final era is the focus of this study.

A number of post-reionization 21~cm experiments that aim to detect the signal and, in some cases, map this signal over large fractions of the cosmic volume are already underway (e.g. LOFAR, CHIME, FAST, TianLai, MeerKAT, ASKAP) \citep{LOFAR_2013, CHIME_22, FAST_2019, TIANLAI_14, MeerKAT_2016, ASKAP_2007} --- with several more planned or nearly completed (e.g. PUMA, CHORD, SKA1-MID, DSA-2000) \citep{PUMA_2019, CHORD_2019,  SKAMID_15, DSA2000_2024, HIRAX_16}. Many of these experiments aim to measure the 21~cm auto-power spectrum, a statistic describing the clustering of neutral hydrogen at large scales, by mapping the neutral hydrogen distribution over large redshift volumes and, thus, making accurate measurements of the positions of the baryon acoustic oscillations (BAO) \citep{CosmologyBook, Eisenstein_2005}. 
The BAO are imprints of the sound horizon scale from the early universe and appear as `wiggles' in the 3D power spectrum of large-scale structure. These wiggles are used as a cosmic ruler to constrain the expansion history of the universe, making their measurement the principal goal of post-reionization 21~cm intensity mapping. Any contamination that shifts the location of the BAO peaks and troughs can therefore bias the inferred cosmological parameters. Additionally, because the wiggles are only $\simeq10\%$ of the total power spectrum amplitude, contamination at the part-in-a-thousand level with a similar oscillatory structure can result in percent-level shifts in the BAO features and, hence, the inferred cosmological parameters.

One of the motivations for targeting the 21~cm line for intensity mapping is due to its low rest-frame frequency of $\nu=1.4~$GHz, resulting in the redshifted line having potentially no significant interloping emission lines that would contaminate the observed intensity maps \citep{Kovetz_2017}.\footnote{Line interlopers are also regularly considered in broader intensity mapping work (for instance, higher-$J$ CO rotational lines in CO surveys) \citep{Lidz_2011}.}
Studies have examined emission sources that could be problematic, such as the 18~cm OH line produced by masers \citep{Gong_2011}, where they find the OH contamination can reach 0.1 to 1\% of the total 21~cm power spectrum. However, the OH line biases only the broadband amplitude of the power spectrum; it would not result in significant biases in the positions of the BAO extrema because the contaminating OH emission comes from gas that is distant compared to the gas that emits in 21~cm. 

Another possible cosmological 21~cm interloper is radio recombination lines (RRLs). RRLs are spectral lines produced by the transition between high principal quantum numbers ($n$), and can arise from several different species. In this work, we primarily focus on hydrogen RRLs (HRRL), which originate within galactic $\hii$ regions \citep{EmigThesis} and are expected to dominate the signal. HRRLs have been used as a way to study the physics of these $\hii$ regions directly, with an increasing number of local observations in recent years \citep{Bell_1980, Puxley_1997, Mohan_2002, Kepley_2011}, generally targeting galaxies with high nuclear star formation rate densities, and two observations at appreciable redshifts ($z>0.5$) \citep{Emig_2019, Emig_2023}. We also expect carbon to produce cosmological RRLs, where these carbon radio recombination lines (CRRLs) are produced in the diffuse, predominantly neutral cold interstellar gas, with typical conditions of $T_e =50-100$~K, $n_e \sim 10^{-3}-10^{-1} ~ \rm cm^{-3}$.\footnote{Helium RRLs may also introduce contamination to the 21~cm signal, but it should trace HII regions just like hydrogen but emit with significantly smaller fluxes owing to its lower abundance. Carbon is different in that it resides in colder gas that is more favorable for RRL emission.} The effect of hydrogen RRLs on intensity mapping experiments has been briefly examined in earlier work by \citet{Petrovic_2011} (referred to herein as \citetalias{Petrovic_2011}), who concentrated on the reionization era where the 21~cm brightness temperature is $\sim 100\times$ larger and so subtle effects from RRLs are less important.

The sample of RRL detections at $z\gtrsim0.5$ is rather limited because the brightness temperature and optical depth of a single RRL is extremely low.
However, the distinctive feature of RRL contamination is that many weak transitions can overlap a single 21~cm band, and transitions close in rest-frame frequency to the 21~cm line can produce oscillatory cross-terms rather than only smooth broadband contamination. In this work, we develop models for RRL emission to assess their impact on post-reionization 21~cm intensity mapping surveys. Additionally, we explore the effect RRLs have on cosmological parameter estimates derived from the baryon acoustic oscillations in the 21~cm power spectrum. 

In section \ref{sec:RRL_BrightTemp}, we describe our methodology for estimating the RRL line brightness temperature and introduce our physically-motivated RRL models. In section \ref{sec:PowerSpectra}, we present our formulation for calculating the RRL auto-power spectra, the 21-RRL cross-power spectra, and the RRL-RRL cross-power spectra for hydrogen. We present and discuss our results for the RRL power spectra in section \ref{sec:Results}, including the impact our model predicts for BAO estimates, and conclude in section \ref{sec:Conclusions}. In our calculations, we use a $\Lambda \rm CDM$ cosmology with parameters roughly consistent with current concordance values from Planck of $\Omega_b = 0.0456$, $\Omega_m = 0.27$, $\Omega_\Lambda=0.73$, $n_s = 0.96$, $\sigma_8 = 0.819$, and $h_{100} = 0.7$ \citep{2020_Planck}.

\section{RRL Brightness Temperature}
\label{sec:RRL_BrightTemp}

In this section we estimate the mean brightness temperature for RRL emission from a given redshift, and compare these estimates to the mean brightness temperature for the redshifted 21~cm signal. Our derivation largely follows \citetalias{Petrovic_2011}, and we highlight key differences. RRLs behave as any other electronic transition with quantum number $n$ with a frequency given by
\begin{equation}
    \nu_{\rm RRL} = cR_{\rm M}\left(\frac{1}{n^2}-\frac{1}{(n+1)^2}\right)\Biggr\rfloor_{n\gg1} \approx 1.4247 ~{\rm GHz} \left( \frac{n}{166}\right)^{-3},
\end{equation}
where $R_{\rm M}$ is the reduced-mass corrected Rydberg constant, and the latter expression only applies to RRLs with $\Delta n = 1$. These very high transitions are observed up to $n\sim1000$ and so they follow a picket fence in frequency that we include in our modeling \citep{Stepkin_2007}. For CRRLs, the recombined electron occupies a state very far from the inner carbon nucleus, and so the outer electron mostly interacts with the net $+1$ charge of the $\mathrm C^{+}$ atom. This results in the atomic structure being well described as a hydrogen atom, with the main carbon-specific differences appearing through the gas properties.

In our work, we only consider lines for which the electronic states differ by a principal quantum number of $\Delta n=1$, as these are the strongest transitions. Additionally, we only consider those lines capable of being redshifted into the 21~cm survey range considered here $0<z_{21}<6$ ($200 \rm ~MHz \leq \nu \leq 1.4~ GHz$), corresponding to transitions with $n>86$. As we will see, we find these restrictions do not affect our predictions for the contamination.

The frequency difference between adjacent principal quantum numbers corresponds to a comoving distance between lines that redshift to the same frequency of
\begin{equation}
    \Delta x = \frac{c\,(1+z) \, d\nu}{H(z)\; \nu_{\rm{obs}}},
\end{equation}
where $d\nu$ is the frequency spacing between two distinct RRLs, or alternatively a RRL and the 21~cm line, and $H(z)$ is Hubble's parameter at redshift $z$ (we are ignoring peculiar velocities). Lines at comparable separations to the BAO scale can obstruct precise distance determinations from BAO -- one of the primary goals of post-reionization 21~cm intensity mapping. As an example, the frequency separation between the 21~cm line and the closest RRL ($166\alpha$) is $d\nu = 4.3~$MHz. At a redshift of $z = 1$, the comoving separation between these lines is $\Delta x(z=1) \approx 16.8~$Mpc,  whereas for a redshift of $z = 5$ the separation is $\Delta x(z=5) \approx 9.6$~ Mpc. The next closest RRL ($167\alpha$), $\Delta x(z=1) = 75$~Mpc and $\Delta x(z=5) = 50$~Mpc, are within a factor of a few to the BAO scale of 140 Mpc. 

The frequency bandpass employed for the analysis of a 21~cm intensity mapping survey corresponds to at least a few hundred comoving megaparsec in order to capture the BAO scale -- and so there are typically several RRLs coming from the same structures as the 21~cm emitters that fall within the bandpass. Additionally, there will be a suite of lower redshift structures producing RRL interlopers for which their 21~cm emission does not fall within the survey bandpass, which yields another form of contamination.

As HRRLs are emitted primarily within $\hii$ regions \citep{EmigThesis} and CRRLs are associated with cold neutral gas that feeds star formation, we assume the RRL luminosity largely traces the star formation rate density (SFRD) (\citep{Emig_2019}; \citetalias{Petrovic_2011}). We now estimate the brightness temperature of RRLs for each of these types, starting with the intensity of a source in the emitting frame, 
\begin{equation}
\label{I_em}
    I_\nu =  \frac{1}{4\pi}\int dr {\cal L}_{\rm \nu_{\rm RRL}}^{\rm RRL}~\rho_{\rm SFR}^{\text{phys}}(r)~G\left(\nu - \nu_{\rm RRL}\left[1+\frac{H(z)r}{c}\right]\right),
\end{equation}
where ${\cal L}^{\rm RRL}_{\nu_{\rm RRL}}$ is the RRL luminosity per unit star formation rate for a line with rest-frame frequency $\nu_{\rm RRL}$, $\rho_{\rm SFR}^{\text{phys}}$ is the SFRD at physical coordinate $r$, $\nu$ is the frequency of interest, and $G$ is the line-spread function which can include both the instrumental resolution and intrinsic broadening, normalized so that $\int d\nu\, G = 1$. For the estimates here, we are interested in the average intensity of the Universe in the emitted frame, so we will replace $\rho_{\rm SFR}^{\text{phys}}(r)$ with its spatial average $\overline{\rho}_{\rm SFR}^{\text{phys}}$. If we approximate the line spread function as a Dirac delta function\footnote{This is a reasonable approximation because using a realistic linewidth only results in smoothing on small scales, which is not the focus of 21~cm intensity mapping experiments.}, we can evaluate the integral using the identity $\int dx ~g\left(x\right)\delta^D\left(f\left(x\right)\right) = \frac{g\left(x_0\right)}{\left|f'\left(x_0\right)\right|}$, where $x_0$ is the root of $f(x)$, which simplifies eq.~\eqref{I_em} to
\begin{equation}
    \overline{I}_{\nu} =\frac{c \, {\cal L}^{\rm RRL}_\nu \; \overline{\rho}_{\rm SFR}^{\rm phys}(z)}{4\pi \, H(z)\, \nu}  .
\end{equation}
 The mean observed brightness temperature relates to the mean intensity in the emitted frame via $\overline{T}_b = c^2 \overline{I}_{\nu_{\rm em}} / [2k_b\nu_{\rm em}^2(1+z)]$, which gives
\begin{equation}
\label{brightnesstemp}
    \overline T_b^{{\rm RRL}_n}= \frac{c^3 {\cal L}^{\rm RRL}_{\nu_{\rm obs}\left(1+z\right)}\;\overline{\rho}_{\rm SFR}(z)}{8\pi k_b \;H(z) \;\nu_{\rm obs}^3\left(1+z\right)} ,
\end{equation}
where we have also changed to using the comoving SFRD via $\overline \rho_{\text{SFR}}^{\text{phys}} = \overline \rho_\text{SFR}(1+z)^3$.

Next, we want to relate ${\cal L}^{\rm RRL}$ to the observed emission in RRLs. We start with the specific line luminosity, $L^{\rm RRL}_\nu$, and the specific continuum luminosity of a galaxy, $L^{C}_{\nu}$.  We have 
\begin{equation}
    \label{flux_relation}
    L^{\rm RRL}_\nu =f L^{C}_\nu e^{-\tau_{\rm ff}}\left(e^{-\tau_{\rm RRL}} - 1\right)  ,
\end{equation}
where 
$\tau_{\rm RRL}$ is the typical line optical depth at frequency $\nu$ of a single RRL producing region, $\tau_{\rm ff}$ is the free-free optical depth for the region at frequency $\nu$, and $f$ is the fraction of the continuum that is absorbed by these optical depths (set by the covering fraction of particular RRL regions). Assuming that the optical depth of the line is small ($\tau_{\rm RRL} \ll 1$), our flux relation simplifies to 
\begin{equation}
    \label{specificLineFlux}
    L^{\rm RRL}_\nu \approx -f \tau_{\rm RRL} L^{C}_\nu e^{-\tau_{\rm ff}}  . 
\end{equation}
The synchrotron luminosity of a single star-forming region is \citep{Yun_2001}\footnote{The synchrotron luminosity associated with star formation is larger than the free-free luminosity associated with star formation for the frequencies of interest, $\nu \lesssim 10~$GHz \citepalias{Petrovic_2011}.} 

\begin{equation}
\label{SFR-RadioLum}
    L_\nu^{C} = 2.2 \times10^{28} \text{erg} ~\text{s}^{-1}~\text{Hz}^{-1} \left(\frac{\text{SFR}}{1 \, \text{M}_\mathSun ~\text{yr}^{-1}}\right) \left(\frac{\nu}{1~ \text{GHz}}\right)^{-0.8},
\end{equation}
where $\rm SFR$ is the star formation rate. This scaling arises because the synchrotron-emitting cosmic-ray electrons are accelerated by supernovae, tracing the SFR. By using this relation, we assume that only the synchrotron continuum luminosity scales linearly with SFR, while the RRL optical depth itself is independent of SFR. Galaxies with a lower SFR may have smaller $\hii$ region covering fractions, which would cause the RRL luminosity to scale stronger than linearly with SFR. Because our optical depth models are supported by highly star-forming galaxies, using this linear SFR scaling may overestimate the RRL emission from galaxies with a lower SFR. We can integrate the specific line luminosity (eq.~\eqref{specificLineFlux}) to get a total line luminosity, yielding 
\begin{equation}
\label{RRLLum}
	L^{\rm RRL} =  -f e^{-\tau_{\rm ff}}\left(\int d\nu  \tau_{\rm RRL}(\nu)\; L_\nu^{C} \right)   \approx  -f \tau_{\rm RRL} L_{\nu_{\rm RRL}}^{C}e^{-\tau_{\rm ff}} \times \frac{\nu_{\rm RRL} \, \Delta V}{c}  ,
\end{equation}
where for the last equality we have defined $\Delta V$ to be the effective velocity width of the line, set by the Doppler and velocity broadening of RRL-producing regions and also their velocity dispersion in the galaxy.

In this work, the optical depth is a signed quantity where a negative $\tau_{\rm RRL}$ corresponds to stimulated emission, while positive $\tau_{\rm RRL}$ corresponds to continuum absorption. In the following brightness temperature estimates, we quote
the amplitude of the RRL contribution. Equivalently, the factor $\tau_{\rm RRL}$
appearing in eqs \eqref{eqn:TbRRL}--\eqref{eqn:TRRL_Tb2} denotes the magnitude of the line optical depth, rather than its signed value. The sign of the RRL contribution is retained separately when computing the cross-power spectra.

Combining eqs.~\eqref{brightnesstemp}, \eqref{SFR-RadioLum}, and \eqref{RRLLum}, and using the definition ${\cal L}^{\rm RRL} = L^{\rm RRL}/{\rm SFR}$, we obtain our final expression for the brightness temperature of a single RRL with quantum number $n$, rest frame frequency $\nu_{\rm RRL}$, and emitted at redshift $z_{{\rm RRL} n}$:
\begin{eqnarray}
    \overline T_b^{{\rm RRL}_n}=\frac{2.2 \times10^{28} \text{erg} ~\text{s}^{-1}~\text{Hz}^{-1}}{8\pi k_b H(z_{{\rm RRL} n})} \frac{c^2}{\nu_{\rm obs}^{2}(1+z_{{\rm RRL} n})^{0.8}}\left(\frac{\overline \rho_{\rm SFR}}{1 \text{M}_\mathSun~\text{yr}^{-1}}\right) \nonumber
    \\\times\left(\frac{\nu_{\rm obs}}{1 \; \text{GHz}}\right)^{-0.8} f \tau_{\rm RRL}\Delta V.
    \label{eqn:TbRRL}
\end{eqnarray}
We can combine terms and insert fiducial hydrogen values, recovering the form of the HRRL brightness temperature at a single redshift:\footnote{Our expression for $\overline{T}_b^{{\rm RRL}_n}$ is a correction to the relation in \citetalias{Petrovic_2011}, where, when evaluated at $z=2$, has a coefficient that is a factor of three times larger. Also, our expression's redshift dependence scales with an additional factor of $\left(1+z\right)$.}
\begin{align}
	\overline T_b^{{\rm RRL}_n} = 2.2\times10^{-5} \text{ K}~\frac{\left(1+z_{{\rm RRL} n}\right)^{-0.8}}{\sqrt{\Omega_M\left(1+z_{{\rm RRL} n}\right)^3 + \left(1-\Omega_M\right)}}\left(\frac{\overline \rho_\text{\rm SFR}}{0.1 ~\text{M}_\mathSun ~\text{yr}^{-1} \text{Mpc}^{-3}}\right)\nonumber\\
    \times\left(\frac{f \tau_{\rm RRL}}{0.1}\right)\left(\frac{\Delta V}{20 \text{ km s}^{-1}}\right)\left(\frac{\nu_{\rm obs}}{500~\text{MHz}}\right)^{-2.8} \label{TRRL_Tb}.
\end{align}
Finally, we write down the brightness temperature for any RRL at $z_{{\rm RRL} n}$, for redshifts where the cosmos is matter dominated, with $H(z) \approx H_0 \sqrt{\Omega_M} (1+z)^{3/2}$, as:
\begin{align}
	\overline T_b^{{\rm RRL}_n} \approx 1.8
    \times10^{-6} \text{ K}~&\left(\frac{\overline \rho_{\rm SFR}}{0.1 ~\text{M}_\mathSun ~\text{yr}^{-1} \text{Mpc}^{-3}}\right)\left(\frac{f \tau_{\rm RRL}}{0.1}\right) \nonumber\\
    &\times\left(\frac{\Delta V}{20 \text{ km s}^{-1}}\right)\left(\frac{\nu_{\rm obs}}{500~\text{MHz}}\right)^{-2.8} \left(\frac{1+z_{{\rm RRL} n}}{3}\right)^{-2.3} \label{eqn:TRRL_Tb2},
\end{align}
where we choose the parameters in the denominators as the fiducial value for a standard $\hii$ region, motivated by the typical values used in \citetalias{Petrovic_2011} and RRL observations.\footnote{This is also the fiducial model (other than the observed frequency) in \citetalias{Petrovic_2011}, who find this model overpredicts observations of $f\tau_{\rm RRL}$ by almost an order of magnitude. The larger line widths of $\hii$ regions inferred from RRL observations means that $f\tau_{\rm RRL}~\Delta V$ agrees with the same observations.} 
\begin{figure}[h]
	\centering
	\includegraphics[width=1.0\linewidth]{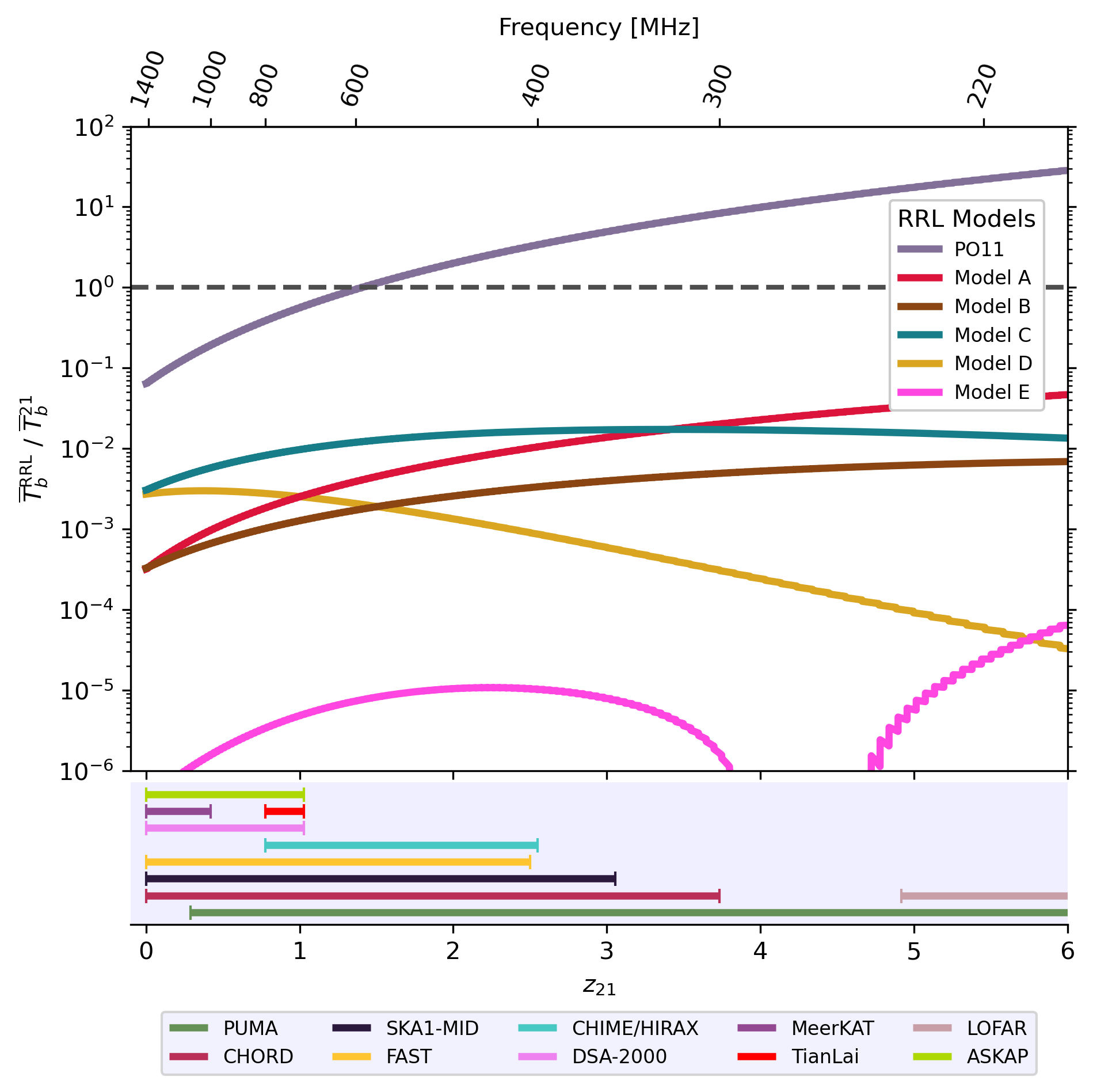}
	\caption{Fractional contamination imparted by all RRLs that overlap with the 21~cm line in frequency space at a specific redshift. Each curve shows the total RRL brightness temperature ratio for different RRL models described in table \ref{tab:RRLModels} (also see section \ref{sec:RRL_BrightTemp} and figure~\ref{fig:tauCompare}), where models A-D are for hydrogen and Model E is for carbon. The previous model in the literature comes from \citetalias{Petrovic_2011} and is shown by the purple curve, where they used a fixed optical depth ($f\tau_{\rm RRL} = 0.1$) as a conservative upper bound for the contamination. The dashed horizontal line marks when $\overline{T}_b^\mathrm{RRL}=\overline{T}_b^{21}$. The total RRL brightness temperature comes both from lines that fall within the 21~cm redshift band, and lines much farther away in physical space that overlap with the observed frequency space. We note these curves should not be extrapolated to cosmic dawn ($z>6$) where the post-reionization 21~cm brightness temperature model in eq. \ref{eqn:T21} no longer holds and the 21~cm signal can be substantially larger. The solid horizontal bars at the bottom of the panel mark the observational bands for different 21~cm intensity mapping experiments.}
	\label{fig:contamSurv}
\end{figure}

As the primary interest of this work is in relation to the 21~cm emission, let us compare the RRL brightness temperature across redshifts with that of 21~cm (e.g. \citep{Ansari_2012}) in the matter-dominated limit, given by


\begin{equation}
\overline T_b^{21} \approx
1.2 \times10^{-4}~{\rm{K}}
\left(\frac{f_{\mathrm{HI}}(z)}{0.01}\right)\left(1+z\right)^{1/2},
\label{eqn:T21}
\end{equation}
where $f_{\mathrm{HI}}(z)$ is the mass fraction of neutral hydrogen mass density to total baryons. For the neutral hydrogen mass density, we use the observationally-consistent results from TNG100 simulations (see the solid line in figure 2 in \cite{Villaescusa-Navarro_2018}).\footnote{We note that these simulations appear to exceed the observed value by $\simeq 2$ at low redshifts ($z\lesssim1$), so we may be overestimating the 21~cm signal.  At higher redshifts, the agreement is better.} For the fiducial parameters in which the parentheses evaluate to unity in eq.~\eqref{eqn:TRRL_Tb2} (e.g., at $z=2$), the single line RRL brightness temperature is smaller than the 21~cm temperature, $\overline{T}_b^\mathrm{RRL}/\overline{T}_b^{21} \simeq 1/6$. At a fixed observed frequency, this ratio decreases away from $z=2$, driven by the RRL redshift scaling at lower redshifts and the frequency scaling at higher redshifts.
    
The RRL brightness temperature signal, however, is composed not only of emission at a fixed redshift, but emission from lines at \emph{any} redshift that overlap in frequency space with the 21~cm signal.  
Figure \ref{fig:contamSurv} shows the level of contamination of the global 21~cm signal, $\overline{ T}_b^{21}$, for several different RRL models, calculated with motivated values of the emitting $\hii$ or diffuse carbon region properties (described below). The gray-purple curve represents the fixed RRL optical depth model used in earlier literature, with $f\tau_{\rm RRL} = 0.1$ and $\Delta V = 20 \rm ~km~s^{-1}$(\citetalias{Petrovic_2011}). This was intended as a conservative upper bound for assessing the RRL contamination during the EoR, rather than a representative physical model. Interestingly, the brightness temperature from RRL interlopers exceeds the 21~cm signal at $z\gtrsim 1$. The high level of contamination from this simplified model motivates the development of more physically-informed RRL models. 

In table \ref{tab:RRLModels}, we describe the physical parameters of the different hydrogen and carbon models we use in our work. For HRRLs, the parameters for these models act as a representative sample of galactic $\hii$ regions, approximately enclosing the upper and lower bounds for HRRL-producing regions. This range is partially motivated by the observed evolution of $\hii$ region density distributions across redshift, inferred from optical line measurements and driven by the evolving pressure of the ISM \cite{Isobe_2023}. For CRRLs, the parameters of Model E are chosen to be approximately consistent with observed CRRL-producing regions \cite{Cros_2025, Salgado_2017, salgado_low-frequency_2017}. The final row gives $f~\rm EM$, which sets the line optical depth normalization. For the free-free attenuation in eqn \ref{eqn:freefreeoptical}, which matters only at high $n$, we take $f=1$, so that $f \rm EM=\rm EM$. For Models A and E, the attenuation does not impact the calculation so the distinction is negligible.

Our chosen emission measures -- which are in part selected to align with some of the RRL observations (which do target galaxies with high nuclear SFR densities) -- likely overestimate the effective path lengths through both diffuse and compact $\hii$ regions. For example, Model A corresponds to a galaxy's (inner few-kpc) synchrotron emission covered by diffuse regions with EM$=10^4\;\mathrm{cm}^{-6}\,\mathrm{pc}$ and line of sight extent $L = \mathrm{EM} / n_e^2 \simeq 100 \rm ~pc$, similar to the thickness of the disk of galaxies and much larger than the path length required to produce the emission measures typical of the diffuse ionized medium in nearby galaxies $\mathrm{EM}\sim1–30\ \mathrm{pc\,cm^{-6}}$\citep{Haffner2009}. Models B and C have $L = f \mathrm{EM} / n_e^2 \simeq (0.1-1)\times f \rm ~pc$, about the size of a dense $\hii$ region if $f=1$ and somewhat above low-redshift observations that suggest a covering fraction of $f=0.3$ of $\hii$ regions for EM$\sim 10^3$ in the inner couple of kpc of star-forming galaxies \citep{MaykerChen_2024}. While their threshold emission measure is substantially lower than the value adopted here, observations indicate that the characteristic \hii-region electron density increases by a factor of $\sim5–10$ from $z\sim0$ to $z\sim2$ \citep{Davies2021} and the SFR surface density increases by roughly an order of magnitude over this interval \citep{Speagle2014,vanderWel2014}. Both trends point toward substantially larger emission measures in high-redshift star-forming galaxies.  
 Model D with $n=10^4$cm$^{-3}$, EM$=10^7\;\mathrm{cm}^{-6}\,\mathrm{pc}$ is even somewhat more extreme. For CRRLs where our representative is Model E, a similar calculus results in a path length through the cold neutral medium of $L\simeq 30~\rm pc$, making use of the carbon-to-hydrogen abundance ratio at solar metallicity.  This path length is guided by Galactic neutral hydrogen emission–absorption studies, which indicate parsec-scale cold neutral medium structures \citep{HeilesTroland2003, Murray2018}.  Simulations further indicate the covering fraction of the cold neutral medium should be high in the interiors of galaxies \citep{Smith2023}.   

The full cosmological RRL signal should thus be interpreted as a superposition of these models, rather than as single models that individually describe all RRL-producing regions. We find that these simplified models --- rather than a complex one that consists of many uncertain quantities --- aid in understanding how the results may change with different assumptions about $\hii$ regions. The curves in figure \ref{fig:contamSurv} show the level of contamination for these models, all of which result in a signal well below the 21~cm signal. However, it is possible that they are still large enough to significantly contaminate the 21~cm intensity mapping and, for hydrogen, the BAO signal.



\begin{figure}[h]
    \centering
    \includegraphics[width=1.0\linewidth]{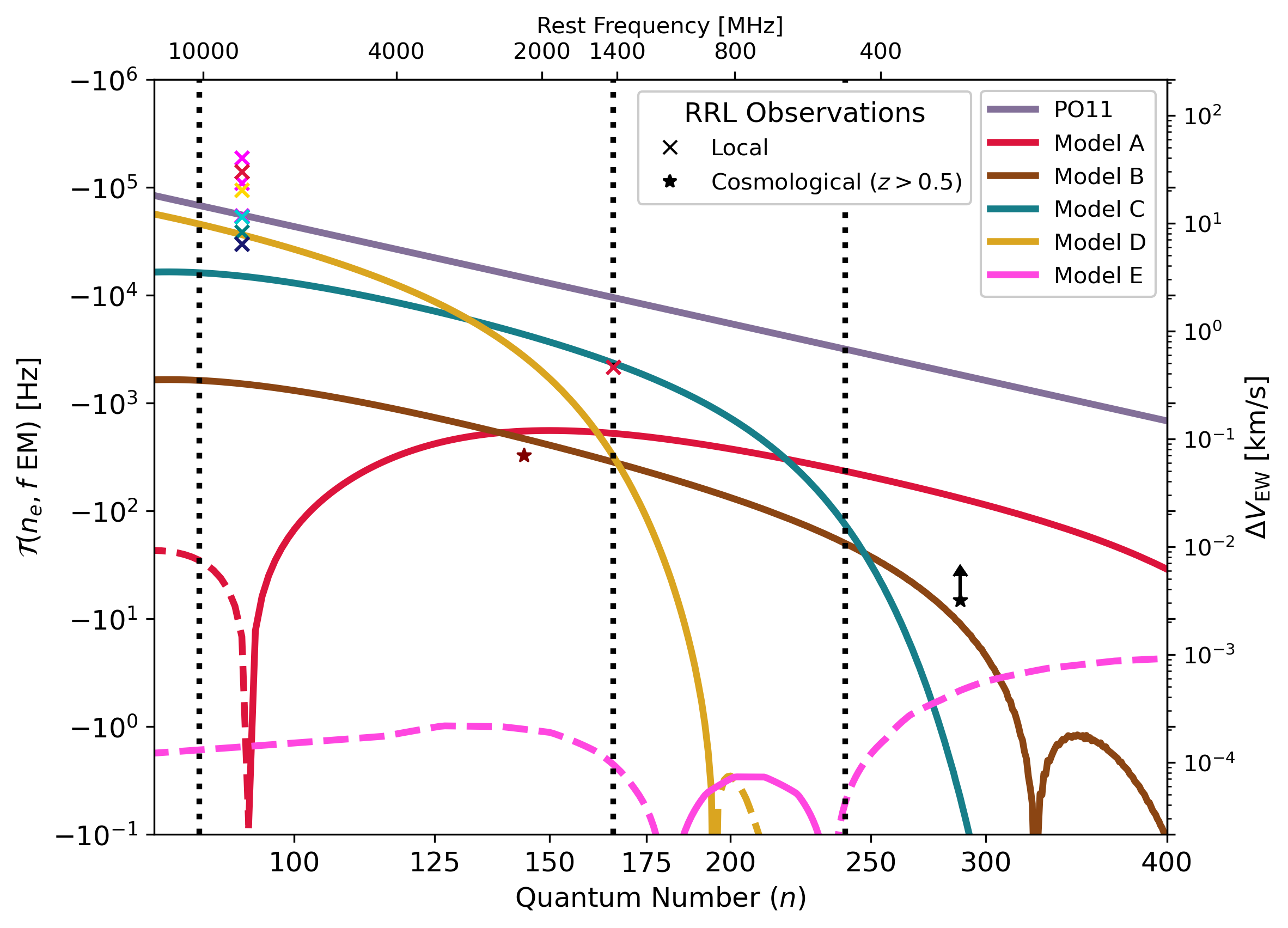}
    \caption{Magnitudes of the cumulative optical depths for the RRL emission models used in this work (table~\ref{tab:RRLModels}), calculated using eq.~\eqref{eqn:optDepthModel}, compared to the fixed optical depth model from \citetalias{Petrovic_2011}. The right-hand y-axis shows the velocity equivalent width ($\Delta V_{\rm EW}$), inferred from the cumulative optical depth. Each marker represents an individual HRRL observation from a star-forming galaxy, and the star symbols mark the two observations at moderate redshifts ($z>0.5$). Dashed segments mark ranges of $n$ where absorption dominates and the cumulative optical depth becomes positive. The three dotted vertical lines mark values of $n$ significant for 21~cm intensity mapping. The line at $n=86$ marks the lowest RRL that can contaminate 21~cm observations over the redshift range considered here  ($0<z_{21} < 6$) through redshifting. The line at $n=166$ corresponds to the RRL whose rest-frame frequency is closest to the 21~cm line ($1420$ MHz). The line at $n=240$ corresponds to the RRL whose rest-frame frequency (and, hence, $z=0$ emission) most closely matches that of 21~cm signal emitted at $z_{21} = 2$.}
    \label{fig:tauCompare}
\end{figure}

To predict $\overline{T}^{{\rm RRL}_n}_b$ using eq.~\eqref{eqn:TRRL_Tb2}, we require both the covering fraction of the $\hii$ and carbon regions in the RRL-emitting galaxy ($f$), and the integrated line optical depth, given by $f \int \tau_{\rm RRL} d\nu $. In this work, we include RRL models motivated by the physical properties of $\hii$ and carbon regions and observations of RRL-emitting regions \citep{Draine_textbook}. Namely, the frequency-integrated optical depth can be related to the properties of the emitting region via 
\begin{equation}
    \label{intTau}
    \int \tau_{\rm RRL}\left(\nu\right) d\nu = 2.046 \text{ MHz} ~\cdot e^{\chi_n} ~\left(\frac{T_e}{\text{K}}\right)^{-5/2}\frac{\rm EM}{\text{cm}^{-6}~\text{pc}}b_n\beta_n,
\end{equation}
where $\chi_n \equiv 1.58\times10^5~\text{K}/(n^2T_e)$, $n$ is the quantum energy level, $T_e$ is the electron temperature, $\rm EM$ is the average emission measure, $b_n$ describes the departure of the gas from local thermodynamic equilibrium, and $\beta_n$ is the stimulated emission correction factor \citep{SalemBrockelhurst_79}. Determining the departure coefficients $b_n$ and $\beta_n$ requires a detailed modeling of the collisional excitation within the emitting gas; we use the tables of these coefficients provided in \cite{Salgado_2017}.\footnote{While this work addresses carbon lines, we note the tables used in our calculation were provided for hydrogen via private communication.} For carbon, the departure coefficients calculated in \cite{Salgado_2017} did not use sufficient convergence criteria --- as discussed in \cite{Prozesky_18} --- and should therefore be interpreted with caution. Specifically, the differences may either come from the dependence of the convergence rate on physical conditions, or intrinsic variability in the departure coefficients themselves. We therefore include only a single representative carbon model and do not interpret its amplitude as an overly precise prediction.
These tables assume Milky Way-like values for the continuum intensity, such as the Galactic synchrotron emission. However, such continuum emission is important at lower-$n$ than what is relevant for this study, as at the most relevant $n$ collisions shape the level populations \cite{Prozesky_18}. 


\begin{table}
    \centering
    \begin{tabular}{lcccccc}
        \toprule
        & \multicolumn{5}{c}{Hydrogen}
        & \multicolumn{1}{c}{Carbon} \\
        \cmidrule(lr){2-6}
        \cmidrule(lr){7-7}
        Model & PO11 & A & B & C & D & E \\
        \midrule
        $T_e\;(\mathrm{K})$ 
        & -- & 8000 &  8000 & 8000 & 8000 & 100\\
        
        $n_e\;(\mathrm{cm}^{-3})$ 
        & -- & 10 & $10^3$ & $10^3$ & $10^4$ & $0.05$\\
        
        $f~\mathrm{EM}\;(\mathrm{cm}^{-6}\,\mathrm{pc})$ 
        & -- & $10^4$ & $10^5$ & $10^6$ & $10^7$  & 0.02\\
        \bottomrule
    \end{tabular}
    \caption{Parameters for the RRL models used in this work in eq.~\eqref{eqn:optDepthModel}. The \citetalias{Petrovic_2011} model is the previous model from the literature, with a fixed RRL optical depth and effective velocity width at all redshifts ($\tau_{\rm RRL} = 0.1$ and $\Delta V = 20 \rm ~km~s^{-1}$). Both electron density ($n_e$) and emission measure ($f ~\rm EM$) are values consistent with those found in the literature for $\hii$ regions \citep{Draine_textbook} and RRL-producing carbon regions \citep{Salgado_2017, Cros_2025}.}
    \label{tab:RRLModels}
\end{table}

We define the cumulative optical depth for each line as the combination of the integrated line optical depth and the continuum optical depth for the RRL producing region. This extra term results from the attenuation by free-free absorption of the RRL emission and so our cumulative RRL optical depth model can be written as
\begin{equation}
    \mathcal{T} \equiv e^{-\tau_{\rm ff}}\int\tau_{\rm RRL}(\nu)d\nu,
    \label{eqn:optDepthModel}
\end{equation}
where the free-free optical depth $\tau_{\rm ff}$ is given in \cite{BlueBook} by
\begin{equation}
    \tau_{\rm ff} = 0.08235~\left(\frac{T_e}{\rm K}\right)^{-1.35}\left(\frac{\nu_{\rm RRL}}{\rm GHz}\right)^{-2.1}\left(\frac{\rm EM}{\rm cm^{-6}~pc}\right)~.
    \label{eqn:freefreeoptical}
\end{equation}
With this convention, $\mathcal{T}<0$ corresponds to stimulated emission, while
$\mathcal{T}>0$ corresponds to absorption against the continuum. The contribution of the continuum optical depth in figure \ref{fig:tauCompare} strongly suppresses the cumulative optical depth at high $n$ (corresponding to lower observed frequencies). The absence of this term results in much stronger predicted contamination at high redshifts.

To test whether these hydrogen RRL models are consistent with observations, we compare them with measurements of extragalactic HRRL emission in figure \ref{fig:tauCompare}, where we plot eq.~\eqref{eqn:optDepthModel} with a fixed electron temperature of $T_e = 8000$~K as this matches the temperature of $\hii$ regions over a broad range of metallicity 
\cite{Draine_textbook}. Observations of HRRLs show a larger line velocity width with values of $100\lesssim\Delta V\lesssim 360 \rm ~km ~s^{-1}$ \citep{Manti_2015} than the fiducial values in eq.~\ref{eqn:TbRRL} and in \citepalias{Petrovic_2011}. However, our physically motivated models depend only on the frequency-integrated optical depth and therefore do not directly depend on the linewidth. \footnote{Our model may be further optimistic (although we expect only minimally) as the synchrotron continuum we use in our model is from cosmic ray electrons, which may be diffuse in these lower frequencies and so lead to a lower covering fraction $f$.}

Additionally, the strongest RRL contamination is present at high quantum numbers $n\gtrsim 90$, where the cumulative optical depth is negative from the masing effect in the $\hii$ region. Observations of RRLs most often report the full width at half maximum and the peak line and continuum flux densities of the region, and so we assume a Gaussian line profile for each observed RRL to calculate an approximate integrated optical depth for the line. Figure \ref{fig:tauCompare} shows selected HRRL observations converted to the appropriate units (cross markers), and compares them to our models (solid curves). Each measurement is from starburst galaxies in the local universe ($z<0.02$) \citep{Anantharamaiah_1993, Mohan_2001, Phookun_1998, Roy_2005, Zhao_1996}, except for the two star points, which represent the only two moderate redshift ($z>0.5$) observations \citep{Manti_2015, Emig_2019, Emig_2023}.\footnote{For the measurements in \citet{Emig_2019, Emig_2023}, the reported values are already of the form of eq.~\eqref{intTau}, and so we simply report their values.} The \citetalias{Petrovic_2011} model is consistent with low-$n$ RRL observations ($n=92$), but is one order of magnitude larger than the measurements at $n = 150$ and $170$. The agreement at low $n$, even with the much larger optical depth, results from their narrow assumed linewidth of $\Delta V = 20 \rm ~km~s^{-1}$. This discrepancy motivates our more physical RRL models in figure \ref{fig:tauCompare}, which we use in the remainder of the paper as a comparison to \citetalias{Petrovic_2011}.

Model E predicts a smaller RRL brightness temperature contribution to the 21~cm signal than our hydrogen models, as shown in figure \ref{fig:contamSurv}. This comparison suggests that for physical properties producing the majority of the lines, CRRLs produce a subdominant contribution to the total RRL contamination. We therefore focus the following sections only on HRRLs.
\section{Power Spectra}\label{sec:PowerSpectra}
While the brightness temperature of RRLs offers an intuitive quantification of the total contamination level, it does not directly translate to the contamination of the principal 21~cm observable in 21~cm intensity mapping experiments, the power spectrum. RRL emission produces three distinct types of contaminating power spectra; the RRL auto-power spectra ($P_{\mathrm{RRL}}$) produced through the correlation of each RRL with itself, the 21~cm-RRL cross-power spectra ($P_{\mathrm{21-RRL}}$) arising from the correlation of HRRL emission from ionized regions at the same redshift as the 21~cm emission of interest, and the RRL-RRL cross-power spectra ($P_{\mathrm{RRL-RRL}}$) produced by the correlation of each RRL with a different RRL. In this section, we extend our estimates from brightness temperature to the contamination of the 21~cm auto-power spectra, and develop the formalism for the RRL auto-power, 21~cm-RRL cross-power, and RRL-RRL cross-power spectra.

In the remainder of this work, we restrict our power spectrum calculations to HRRLs, which provide the dominant contribution for the models considered in section \ref{sec:RRL_BrightTemp}. However, the power spectrum formalism described below is also applicable to CRRLs, although with different bias factors and shot noise prescriptions. 
\subsection{Temperature Fluctuations}
We begin with the picture that there is contamination to the observed 21~cm brightness temperature by RRLs, and the total observed brightness temperature in direction $\nhat$ observed at frequency $\nu$ is given by
\begin{equation}
    T(\nu, \nhat) = T_b^{{\rm 21}}(\nu_{21}, \nhat) + \sum_n T_b^{{\rm RRL}_n}(\nu_{{\rm RRL} n},  \nhat),
\end{equation}
where  $\nu_{21}$ is the emitted 21~cm frequency ($1.4\rm~GHz$), $\nu_{{\rm RRL} n}$ is the emitted frequency of a particular line, $1+ z_{21} = \nu_{21}/\nu$ and $1+ z_{{\rm RRL} n} = \nu_{{\rm RRL} n}/\nu$, the summation is over all RRLs, and temperatures have been scaled to their redshift zero values.


In this paper, we focus on the fluctuations in the brightness temperature about the all-sky averaged mean.  We define the overdensity in the 21~cm brightness temperature as $\delta_{21} \equiv  T_b^{{\rm 21}}(z_{21}, \nhat)/ \overline T_b^{{\rm 21}}(z_{21}) - 1$, where $\overline{T}_b^{21}$ is the mean brightness temperature, and we define an overdensity in the brightness temperature of the RRL from $n$ similarly and call it $\delta_{{\rm RRL},n}$.  The observed temperature fluctuations about the mean temperature can be written in terms of these overdensities as
\begin{equation}
    \delta \overline{T}(\bfx_{21}) = \overline{T}_{21}(z_{21})\delta_{21}(\bfx_{21}) +  \sum_{n} \overline{T}_{{\rm RRL} n} (z_{{\rm RRL} n})\, \delta_{{\rm RRL},n} (\bfx_{21}+ \Delta \bfx_{n}),
    \label{eqn:Tfluc}
\end{equation}
where $\bfx$ is comoving position, which is often a more convenient coordinate than $\{z_{21}, \nhat\}$.  As we will be considering contamination in the context of 21~cm surveys, eq.~\eqref{eqn:Tfluc} expresses $\delta \overline{T}(\bfx_{21})$ in terms of the comoving position of the 21~cm emission $\bfx_{21}$.  RRLs that fall at the same frequency and direction appear like 21~cm emission at that position, but really are at position $\bfx_{21}+ \Delta \bfx_{n}$ where 
\begin{equation}
    \Delta \bfx_{n} = \int^{z_{{\rm RRL} n}}_{z_{21}} \frac{c}{H(z')}dz',
\end{equation}
where $z_{{\rm RRL} n} \equiv \frac{\nu_{{\rm RRL},n}^\mathrm{emit}\left(1+z_{21}\right)}{\nu_{21}}-1$.
Note that, while the line-of-sight comoving position is redundant with the redshift, we choose to write the mean temperature as a function of redshift since it only depends on the line-of-sight coordinate.



\subsection{Power Spectrum}
The primary statistic used by intensity mapping efforts is the power spectrum.  The contamination by interlopers was investigated in \cite{Lidz_2016}, and we follow their notation, except that we add the additional source of contamination that arises from the cross correlation of different lines. The total power spectrum measured is given by the 21~cm signal and the contamination from RRLs: 
\begin{align}
    P_{\rm  tot}(\bfk)= P_{21}(\bfk) +\overbrace{ \sum_n
         P_{21 \times {\rm RRL}_n}(k)}^{\equiv P_{21-\rm RRL}}
       + \overbrace{\sum_{n,m}
        \frac{1}{\alpha_\perp^2 \alpha_\parallel}P_{{\rm RRL}_n \times {\rm RRL}_m}\left( \frac{k_\perp}{\alpha_\perp} , \frac{k_\parallel}{\alpha_\parallel}\right)}^{\equiv ~ P_{\rm RRL}(n=m) ~+~ P_{\rm RRL-RRL}(n\neq m)}
    \label{eqn:Pkdecomp}
\end{align}
where we have defined
$$\alpha_\perp = \frac{x_{{\rm RRL}_n}}{x_{21}}\quad \mathrm{and}\quad \alpha_\parallel = \frac{1+z_{{\rm RRL} n}}{1+z_{21}} \frac{H(z_{21})}{H(z_{{\rm RRL}_n})}.$$
The $\alpha_\perp$ and $\alpha_\parallel$ map from the actual wavenumber an observer interprets the power spectrum as being observed at to the correct wavenumber for the distance of the RRL. The $n$, $m$, where each sum is over all RRLs that contaminate the survey, includes both auto-correlations when $n=m$, and cross correlations when $n\neq m$.  Cross correlations were not considered in \cite{Lidz_2016} and so eq.~\eqref{eqn:Pkdecomp} is a generalization.  Because cross correlations are most significant when the distance between structures is smaller than the line-of-sight size of the survey (there is a factor we will get to in these power spectra that results in a large suppression in the opposite limit), this allows us to use one set of $\alpha_\parallel$ and $\alpha_\perp$ in eq.~(\ref{eqn:Pkdecomp})  (as they are equal in this limit).  Both are unity in the case where the cross correlation is between a RRL and the 21cm transition.   

For 21~cm, we can write the auto-power as
\begin{equation}
    P_{21}(\bfk) =  \left(\overline T_b^{{\rm 21}}\right)^2 \left[ (b_{21} + f_{\rm RSD} \mu^2)^2 P_{\text{matter}}(k, z_{\rm 21}) + \bar n_{21}^{-1} \right],
        \label{eqn:P21}
\end{equation}
where $b_{21}$ is the linear bias computed following the approach described in \citet{cosmicvisions_2019} of interpolating between IllustrisTNG at $z<2$ and their halo model at $z>2$, $f_{\rm RSD} \mu^2$ is the Kaiser factor that describes redshift space distortions where $f_{\rm RSD} \approx \Omega_m(z)^{0.6}$ and $\mu = k_\parallel/k$, $P_{\text{matter}}(k, z)$ is the linear  matter power spectrum, 
  and $\bar n_{21}^{-1}$ the shot noise. We take the 21~cm shot noise to be negligible ($\bar n_{21}^{-1}=0$) for this preliminary work as 21~cm emitters tend to be more numerous than the star-forming galaxies that should dominate RRL emission. We are most interested in BAO scales, where this shot noise is absolutely negligible.  We do include the RRL shot noise as described below.
  
We can write a similar expression for the RRL auto power spectrum
\begin{equation}
    P_{{\rm RRL}_n \times {\rm RRL}_n}(\bfk) =  \left(\overline T_b^{{\rm RRL}_n}\right)^2 \left[ (b_{{\rm RRL}_n} + f_{\rm RSD} \mu^2)^2 P_{\text{matter}}(k, z_{{\rm RRL}_n}) + \bar n_{{\rm RRL}_n}^{-1}\right],
    \label{eqn:PRRLauto}
\end{equation}
where the redshift space distortions are defined similarly. The bias of RRLs in our model in which the emission is proportional to the SFR should be the same as that of any other line that is also proportional to SFR, allowing us to use the linear CO bias given in \cite{Moradinezhad_Dizgah_2022}. The shot noise is derived by using Schechter function fits to the star formation rates of galaxies -- using that RRLs trace star formation -- with the normalization factors taken from \cite{Smit_2012}.
Adopting fixed Schechter parameters, $\phi^* = 10^{-3} ~\rm Mpc^{-3}$, $\psi^* = 200 \rm ~M_\mathSun ~\rm yr^{-1}$, and $\alpha = -1.6$ \citep{Smit_2012} where $\phi\left(\psi\right) = \phi^* (\psi/\psi^*)^{\alpha} \exp(-\psi/\psi^*)/\psi^*$ and using 
\begin{equation}
    \bar{n}_{\rm RRL_n}^{-1} = \frac{\int d\psi~\phi\left(\psi\right)\psi^2}{\Big[\int d\psi~\phi\left(\psi\right)\psi^2\Big]^2},
\end{equation}
we obtain $\bar{n}_{\rm RRL_n}^{-1} \simeq 180 \rm ~Mpc^3$, which we take to be constant with redshift and the same for all RRL transitions. This value is significantly higher than estimates for the 21~cm signal's shot noise \cite{cosmicvisions_2019}, reflecting that the SFR-tracing signal is weighted toward galaxies at the characteristic $\psi^*$ scale, and so the shot noise is skewed towards the bright end of the SFR function compared to the 21~cm signal.

The cross power spectra are more complex to compute. Let us consider correlating the Fourier modes of a translated field relative to another by the line of sight distance $\Delta x$.  Here $\Delta x$ is the difference between the two sightlines.  Translation results in a phase $\exp[-i k_\parallel \Delta x] $ so we would expect the power is shifted by a phase relative to the previous. However, there is also a finite bandwidth for 21~cm observations, which we parameterize using the fractional redshift width as $f_{\rm BW} \equiv \Delta z /\left(1+z\right)= \Delta \nu / \nu_{\rm obs}$. Now a finite bandwidth results in the observed power spectrum in the volume being a convolution between the volume's transform and the true power spectrum in Fourier space. We define the width of the tophat as $L = c\left(1+z\right)f_{\rm BW} / H(z)$. The tophat observation from $-L/2$ to $L/2$ transforms to a sinc function  $ L \;{\rm sinc}( k_\parallel L/2)$, times another sinc factor for the second overdensity (approximating both as covering the same distance $L$, which holds to the extent the Hubble expansion is constant for the two slightly shifted fields). Therefore, the cross power spectrum is the convolution of this survey window function with the phase-rotated cross power spectrum:
\begin{eqnarray}
{P}_{21\times {\rm RRL}_n} &= & 2\Re \left[ P_{\rm cross}(\mathbf{k}_\perp, k_\parallel) e^{-i k_\parallel   \Delta x} \star  L~{\rm sinc} \left( \frac{ k_\parallel L }{2} \right)^2 \right] \times W_{\Delta x_{21-{\rm RRL}_n}}(\bfk); 
\label{eqn:cross}\\
P_{\rm cross} &\equiv& \overline T_b^{{\rm 21}} \overline T_b^{{\rm RRL}_n}  \left[(b_{21} + f_{\rm RSD} \mu^2)  (b_{{\rm RRL}_n} + f_{\rm RSD} \mu^2) P_{\rm matter}(k) + \bar n_{21, {\rm RRL}_n}^{-1} \right],
\end{eqnarray}
where the star denotes a convolution and the real value owes to the fact that both $\delta_{21}\delta_{{\rm RRL}_n}^*$ and $\delta_{21}^*\delta_{{\rm RRL}_n}$ contribute to the total power. Here, we set the cross-term shot noise to be the same as our RRL auto-term shot noise, $\bar n_{21, {\rm RRL}_n}^{-1}=\bar n_{ {\rm RRL}_n}^{-1}=180 \rm ~Mpc^3$. This is motivated by the fact that the population contributing to the 21-RRL cross shot noise is comparable to that of the RRL-auto contributors. We find that varying or omitting this shot noise term does not significantly affect our results, particularly on the inferred locations of the BAO extrema (as discussed in section \ref{sec:cosmology}).

Note that the same sinc-squared convolved with the unwindowed power-spectrum should also appear in our auto-power spectra due to finite bandwidth effects (e.g. eqs.~\eqref{eqn:P21} and \ref{eqn:PRRLauto}), but we ignore it as it only slightly suppresses the power in modes on the scale of $L$. For the cross power, because of the oscillatory phase and the potential large separations $\Delta x_{21-{\rm RRL}_n}$, the survey window is more important to include. We have not yet specified the function $W_{\Delta x}(\bfk)$, which we will do shortly.

Eq.~\ref{eqn:cross} also has an intuitive limiting case, corresponding to a signal from a pure shot noise contribution. In the case where the power spectrum is scale-independent (i.e., there are only local correlations) the convolution reduces to multiplying by a triangle window function such that
\begin{equation}
    {P}_{21\times {\rm RRL}_n} =  2P_{\rm cross} \exp[-i k_\parallel \Delta x](L-|\Delta x|)/L
\end{equation}
if $|\Delta x| < L$ and zero otherwise. This is exactly as expected, as the local correlations only appear when 21~cm and RRL signals arise from the same structure and fall in the survey bandpass. We have verified numerically this limiting case yields nearly identical results to the full convolution in eq. \ref{eqn:cross}. Consistent with this, figure \ref{fig:PSPLOT_BW} shows only a weak dependence on the observational bandwidth.

Finally we must define the function $W_{\Delta x}(\bfk)$ in eq.~\eqref{eqn:cross}.  Only if the perpendicular wavevector from both lines falls in the same Fourier bin will there be correlations between the 21~cm and RRLs.  Modes will become uncorrelated once $k_\perp x_{21}  - k_\perp x_{{\rm RRL}_n} \gtrsim \pi/\theta_b$, where $\theta_b$ is the angular size of the telescope's beam,
where this condition becomes $k_\perp \Delta x_{21-{\rm RRL}_n} \gtrsim \pi/\theta_b$~.  However, there is a complication that depends on the analysis method.  If the analysis assumes that all structures in the survey are at the same distance from the observer, which is the approximation that is typically made so that each visibility probes a different $k_\perp$, there is the additional condition of approximately $\Delta x_{21-{\rm RRL}_n} \gtrsim L$ for this suppression to occur.  This occurs because when the RRL emission in a survey overlaps spatially with the 21~cm emission, the emission from a given structure in the overlap volume will map to the same wavevectors in both (since the angular separation of the structure in both the 21~cm and RRL is the same).\footnote{An alternative analysis would remap all frequencies to different distances (assuming the emission is 21~cm) and, then, Fourier transform the remapped field. This approach eliminates this condition for suppression as the same structures appear at different physical separations for the RRL emission since it occurs at a different frequency.} Thus, assuming a Gaussian beam for the radio telescope, we expect we can approximate this suppression as:
\begin{equation}
W_{\Delta x}^\perp(\bfk_\perp) \approx \exp\left[- \frac{(\theta_b k_\perp \Delta x)^2}{2\pi}\right] \text {if $\Delta x > L$}
\end{equation}
and unity otherwise.  Taking the example of CHORD for which $\theta_b \approx 5^\circ$, this results in a suppression of modes with $k_\perp\gtrsim 0.1~$Mpc$^{-1}$ for $\Delta x_{21-{\rm RRL}_n} =1000~$Mpc and $k_\perp\gtrsim 1~$Mpc$^{-1}$ for a more relevant value of $\Delta x_{21-{\rm RRL}_n} =100~$Mpc, which is larger than the wavenumbers relevant for the BAO. There is also a parallel suppression from the same effect where parallel Fourier modes will not be binned in the same bin in a manner described by
\begin{equation}
W_{\Delta x}^\parallel(\bfk_\parallel) \approx {\rm rect} \left[\frac{(1-\alpha_\parallel) k_\parallel}{2\pi/L} \right] \text {if $\Delta x > L$},
\end{equation}
and one otherwise. Analogously to $W_{\Delta x}^\perp$, the \emph{if} condition assumes that distance and frequency are treated as linear over the survey in the analysis; correcting this assumption would result in the rect expression holding at all $\Delta x$. The full function is then $W_{\Delta x}(\bfk) \equiv W_{\Delta x}^\perp W_{\Delta x}^\parallel$. In the calculations below, we set $W_{\Delta x}(\bfk)=1$ since the cross-power is dominated by pairs close to each other ($\Delta x < L$, where $W_{\Delta x}(\bfk)\equiv1$) and lines with greater separation are suppressed by the survey window.

Finally, the RRL-RRL power is defined analogously
\begin{eqnarray}
{P}_{ {\rm RRL}_n \times {\rm RRL}_m} &= & \Re \left[ {\cal P}_{\rm RRL-RRL}(\mathbf{k}_\perp, k_\parallel) e^{-i k_\parallel   \Delta x} \star  L~{\rm sinc} \left( \frac{ k_\parallel L }{2} \right)^2 \right] \times W_{\Delta x_{{\rm RRL}_n-{\rm RRL}_m}}(\bfk); 
\label{eqn:cross2}\\
{\cal P}_{\rm RRL-RRL} &\equiv& \overline T_b^{{\rm RRL}_n} \overline T_b^{{\rm RRL}_m}  \left[(b_{{\rm RRL}_n} + f_{\rm RSD} \mu^2)  (b_{{\rm RRL}_m} + f_{\rm RSD} \mu^2) P_{\rm matter}(k) + \bar n_{{\rm RRL}_n, {\rm RRL}_m}^{-1} \right]. \nonumber
\end{eqnarray}

\section{Results}\label{sec:Results}
In this section, we examine the contamination that RRLs impart on the 21~cm auto-power spectrum, and discuss the estimated effect they  have on cosmological measurements. 

\subsection{RRL Power Spectra}
\begin{figure}
    \centering
    \includegraphics[width=1.0\linewidth]{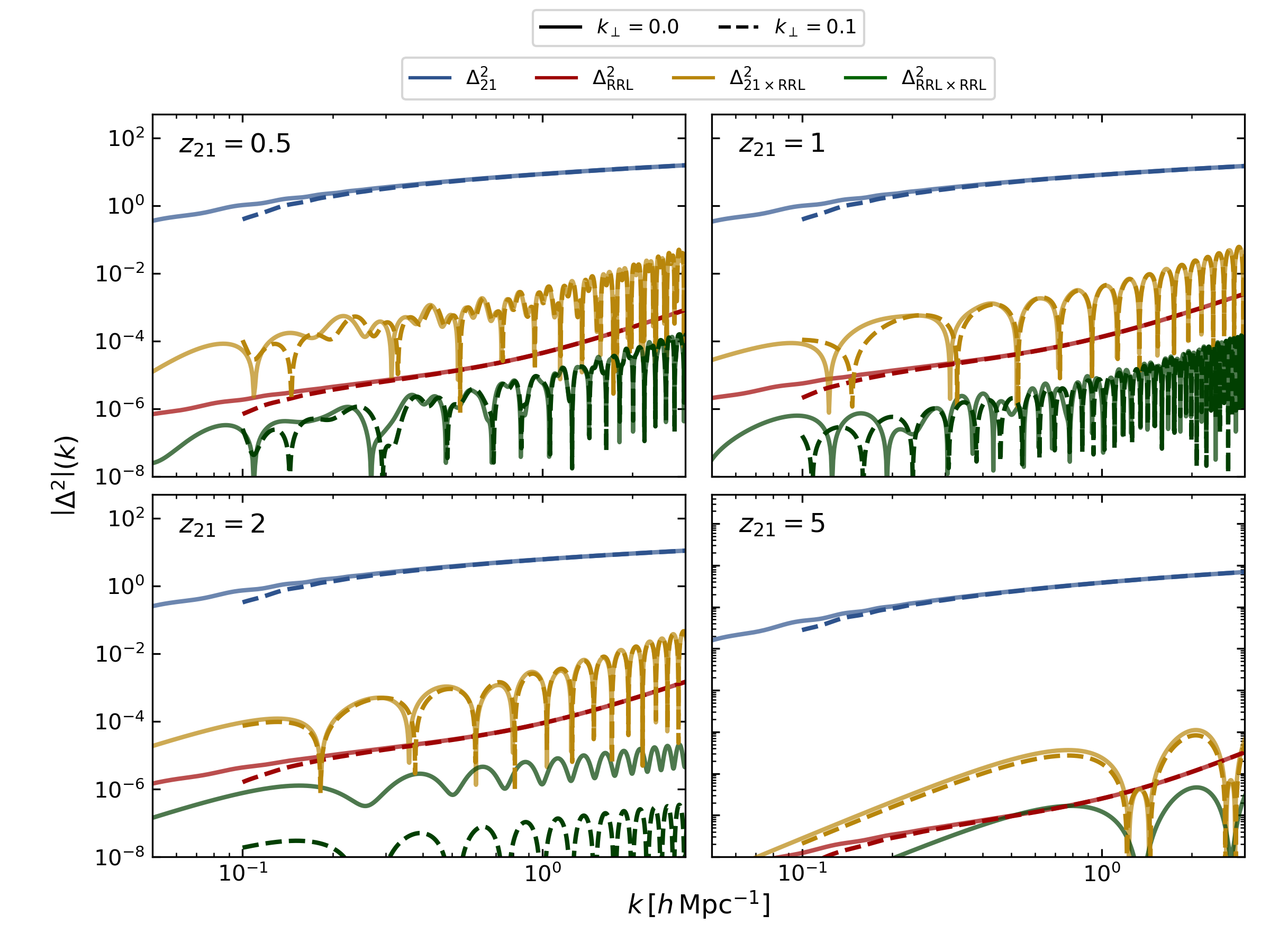}
    \caption{Magnitude of the dimensionless power spectra (eq. \ref{eqn:dimPS}) for each of the three RRL power spectra contributions that contaminate the 21~cm power spectrum (eqns. \ref{eqn:PRRLauto}, \ref{eqn:cross}, and \ref{eqn:cross2} in red, yellow, and green respectively), compared to the dimensionless 21~cm auto-power spectra (blue). The solid lines represent the linear $k_\perp=0$ power spectra, while the dashed lines are the corresponding $k_\perp = 0.1 ~h\rm~Mpc^{-1}$ power spectra of the same type. Each panel uses a different value of $z_{21}$, and all curves are calculated for our fiducial Model C and a bandwidth of $f_{\rm BW}=0.05$ corresponding to width of BAO-scale modes in the power spectrum. We find the choice of bandwidth does not significantly impact the result, as we show in figure~\ref{fig:PSPLOT_BW}.}
    \label{fig:PSPLOT_indiv}
\end{figure}
 We choose Model C from table~\ref{tab:RRLModels} (teal curve in figure~\ref{fig:tauCompare}) as our fiducial RRL model, as it provides an intermediate case that is broadly consistent with RRL observations of $\hii$ regions. Later on we consider all other RRL models from table \ref{tab:RRLModels} -- along with the previous model from \citetalias{Petrovic_2011} -- to get a sense for potential uncertainty.  

 Our procedure is as follows. We begin by choosing a 21~cm observing band of interest, and identifying all RRLs (and their corresponding redshifts) that are capable of overlapping in frequency space with the observing band. For each of those lines, we calculate the brightness temperature using eq.~\eqref{eqn:TbRRL}, and finally their auto-power spectra using eq.~\eqref{eqn:PRRLauto}. For the 21-RRL and RRL-RRL cross-power spectra, we follow the same procedure only with their corresponding power spectra relations, eqs. \eqref{eqn:cross} and \eqref{eqn:cross2}. Finally, we find that lines that are separated by more than $10^3$ Mpc have a negligible contribution to the total signal, and so we omit them from the calculation for efficiency. We also make an approximation for lines with $z_{{\rm RRL} n}<0.05$, which correspond to relatively local emission where each galaxy could realistically be masked and its RRL emission removed observationally. Therefore, in order to retain the contribution of these galaxies without an explicit masking model, we set the minimum distance to be half the observing bandwidth (i.e. $x_{\rm RRL} = \max[x_{\rm RRL}, {\rm BW}(z_{\rm 21})/2]$ where ${\rm BW}(z_{\rm 21})$ is the observing bandwidth of the 21~cm experiment at redshift $z_{\rm 21}$) converted to a comoving radial distance.

Figure \ref{fig:PSPLOT_indiv} shows the results for our fiducial model (Model C, other models are shown below) for different 21~cm redshifts and two different values of $k_\perp$. Each panel shows the 21~cm auto-power spectrum (blue curves) and the three components of the dimensionless RRL power spectra, 
\begin{equation}
    \label{eqn:dimPS}
    \Delta_{\rm{X}}^2 \equiv k^3 P_{\rm{X}}(k) \bigg/ 
    \left[2\pi^2 \left(\overline T_b^{{\rm 21}}\right)^2\right]~,
\end{equation}
where $\rm X$ is a placeholder to indicate the particular lines used to calculate the power spectrum. The range of $k$ values shown, $0.05$ to $3~h~\rm Mpc^{-1}$, corresponds to those targeted by 21~cm intensity mapping efforts, with the lower values of $k\sim 0.1~h~\rm Mpc^{-1}$ aligning closely with the BAO scale that is the principal focus of these 21~cm efforts \citep{Ansari_2012}.

As an example, let us focus on the upper left panel in figure \ref{fig:PSPLOT_indiv} that features the case $z_{21}=0.5$, although the calculations in the other panels are qualitatively similar. The curves showing $\Delta^2_{\rm RRL}$ (red) describe the clustering of all summed RRLs with themselves, and have a similar shape to the 21~cm auto-power spectrum -- tracing the matter power spectrum -- even though most of the lines contributing to $\Delta^2_{\rm RRL}$ come from lower redshifts (see figure \ref{fig:n_space_contam} below). For the other two power spectra contributions ($\Delta^2_{\rm 21 \times RRL}$ in yellow and $\Delta^2_{\rm RRL \times RRL}$ in green), 
the frequency offset of the contributing lines produces oscillations in the power spectrum, corresponding to Fourier space analog of a displacement along the line of sight. The strongest contaminating signal from RRLs comes from the 21~cm-RRL cross-power spectrum, as seen in the yellow curves. This is because the cross-spectrum scales linearly as $\overline T_b^{\rm RRL}/\overline T_b^{21}$, whereas the RRL auto-power spectrum scales quadratically as $\left(\overline T_b^{\rm RRL}/\overline T_b^{21}\right)^2$ (see eqs.~\eqref{eqn:cross}, \eqref{eqn:cross2}, and \eqref{eqn:TRRL_Tb2}).  
In the case of BAO, these oscillations could pose challenges for precise estimates of cosmological parameters from the observed 21~cm auto-power spectrum \citep{Villaescusa_Navarro_2016}. We discuss the magnitude of this effect in section \ref{sec:cosmology}. 
\begin{figure}
    \centering    \includegraphics[width=1.0\linewidth]{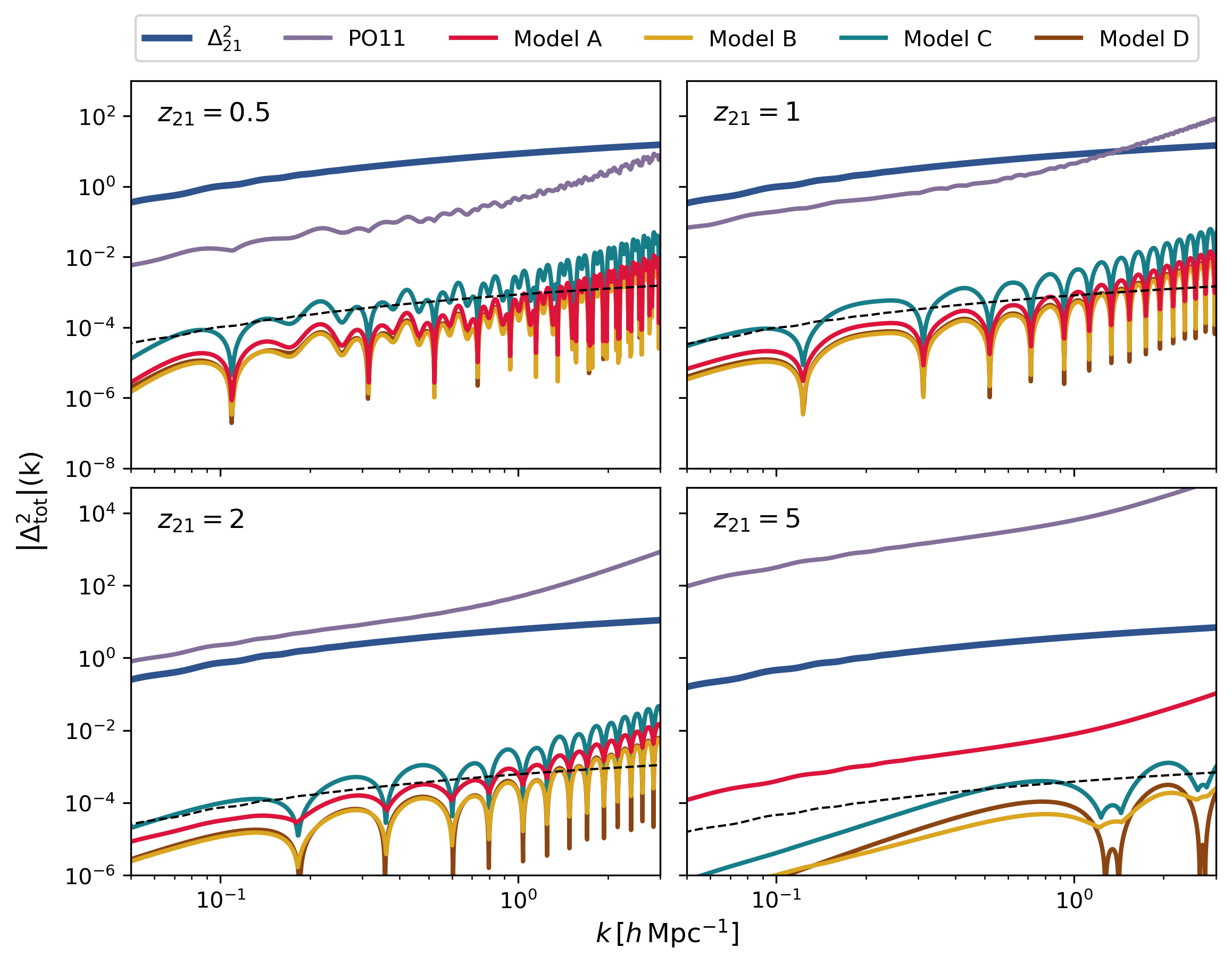}
    \caption{Dimensionless power spectra as in figure \ref{fig:PSPLOT_indiv} showing the 21~cm power spectrum in thick blue and our four models for the HRRL contamination with the thinner curves (see table \ref{tab:RRLModels} and figure~\ref{fig:tauCompare}).  For the latter, the three contaminating RRL power spectra are all summed together ($\Delta^2_\mathrm{contam} = \Delta^2_{\mathrm{RRL}} + \Delta^2_{\mathrm{21 \times RRL}} + \Delta^2_{\mathrm{RRL \times RRL}}$).  We do not show the CRRL model (Model E) as its contamination is much smaller. The redshift in each panel is the redshift of the 21~cm survey, and all are calculated with a bandwidth of $f_{\rm BW}=0.05$ and $k_\perp = 0$. The thin dashed black curve marks $10^{-4}P_{21}$, a level of contamination that may be significant for current or future 21~cm surveys.}
    \label{fig:PSPLOT_resid}
\end{figure}
At $z_{21}=5$ (bottom-right panel of figure~\ref{fig:PSPLOT_indiv}), the cross terms are much smaller than the smooth RRL auto-power spectra terms as the relative size of RRL contamination is now larger.  The RRL lines that contaminate the signal are much farther in the foreground than the 21~cm emission, as we discuss shortly. Because of the geometry of the system, the RRLs that dominate the signal are produced closer to us and, thus the same angle corresponds to a smaller transverse distance than the 21~cm signal. This results in shot noise becoming more significant and dominating the RRL signal at $k\gtrsim 1~h~\rm Mpc^{-1}$. As the 21~cm signal is much farther away, the effect of surface brightness dimming is stronger than on the RRL signal, and so the cross-power spectrum becomes weaker. Additionally, the strong redshift scaling for the RRL brightness temperature at lower redshifts (see eq.~\eqref{eqn:TRRL_Tb2}) results in a weaker signal. The two other redshifts behave as intermediate cases between the $z_{21}=0.5$ and $5$, where strong oscillations appear in the cross-power spectrum but are mildly suppressed by surface brightness dimming.


\begin{figure}
    \centering
    \includegraphics[width=1.0\linewidth]{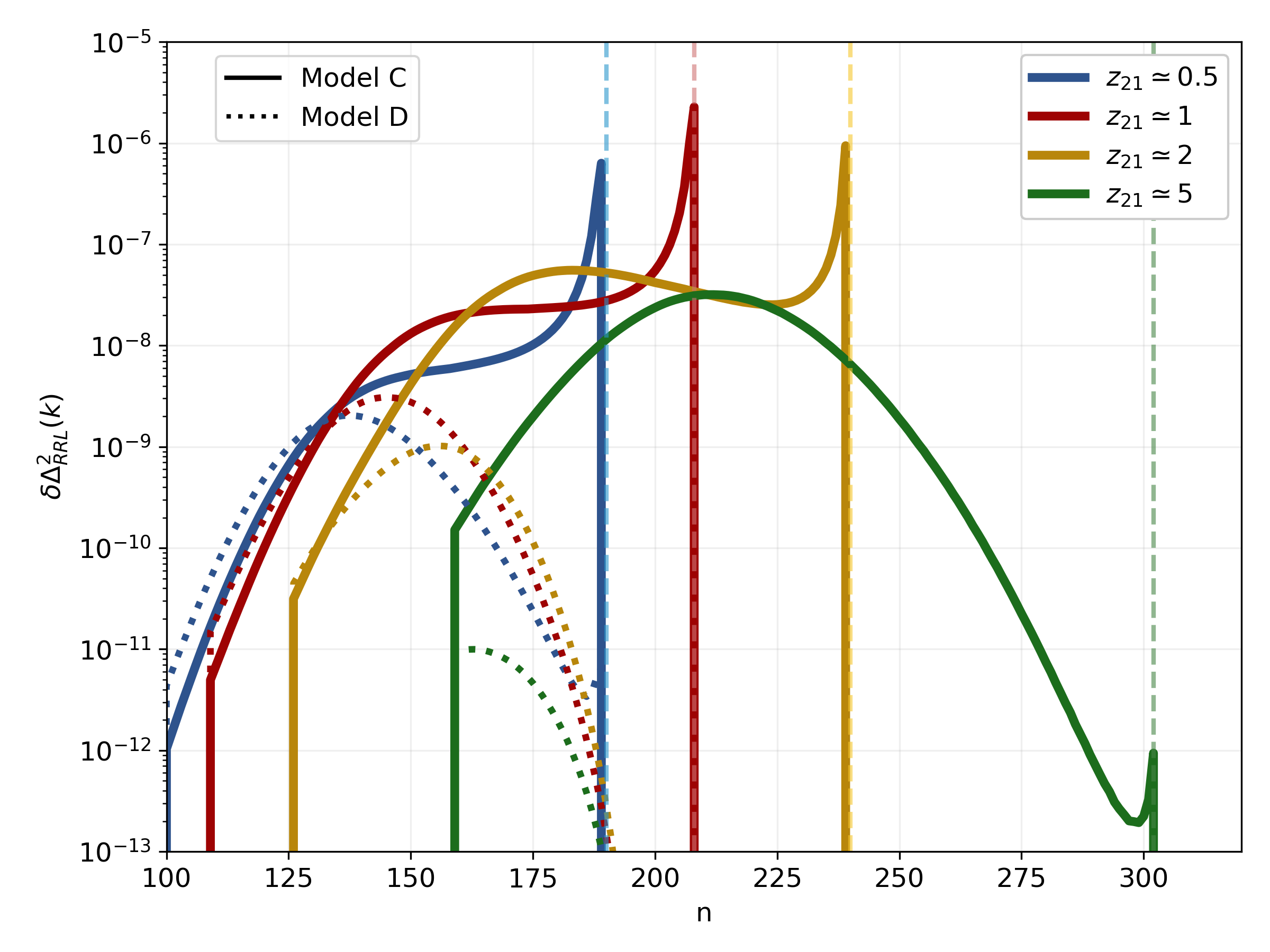}
    \caption{Total power for each RRL ($n$) auto-power spectrum ($\Delta^2_{\rm RRL}$) at $k=0.1~h~\rm{Mpc}^{-1}$ given a 21~cm signal emitted at different redshifts. The solid lines indicate the level of contamination using our fiducial model (Model C), and the dotted lines are for our higher density \hii~region model (Model D). The vertical dashed lines mark for each $z_{21}$ the RRLs with rest-frame frequencies equal to the observed 21~cm frequency, corresponding to $z_{\rm RRL}=0$.}
    \label{fig:n_space_contam}
\end{figure}

Figure \ref{fig:PSPLOT_indiv} also shows the impact of the perpendicular wavenumber ($k_\perp$), where the dashed curves corresponding to $k_\perp=0.1\,h\,{\rm Mpc}^{-1}$ and the solid curves to $k_\perp=0$. We only consider small line-of-sight wavenumbers relevant for 21~cm intensity mapping, where the resulting differences are modest. The non-zero wavenumber results in a mild suppression of the cross-power spectrum --- owing to the shift in the matter power spectrum --- which is most visible in the dashed yellow and green curves at intermediate redshifts.

Next, in figure~\ref{fig:PSPLOT_resid} we present the stacked RRL contamination power spectra ($\Delta^2_\mathrm{contam} = \Delta^2_{\mathrm{RRL}} + \Delta^2_{\mathrm{21 \times RRL}} + \Delta^2_{\mathrm{RRL \times RRL}}$) for all HRRL optical depth models shown in figure \ref{fig:tauCompare}, including the fixed optical depth model from \citetalias{Petrovic_2011}.  We do not show the CRRL model (model E) as its contamination is much smaller. Since the main interest in the context of intensity mapping is how RRLs contaminate precision cosmology measurements, the amplitude and level of oscillations at low wavenumber $k$ of this combined RRL power spectra are particularly important as this is what can distort the inferred BAO signal. Across all redshifts, the fixed optical depth model from \citetalias{Petrovic_2011} predicts significant emission, even exceeding $P_{21}$ at moderate redshifts of $z_{21}\gtrsim 1$. While the observationally-motivated RRL models reduce the stacked power spectrum, the level of contamination is comparable to or exceeds the ($\Delta^2_{\mathrm{contam}}/\Delta^2_{\mathrm{21}}\gtrsim 10^{-4}$) reference level shown by the dashed curve in figure \ref{fig:PSPLOT_resid}. Although this is much smaller than the amplitude of the BAO wiggles themselves, its oscillatory structure could in principle shift their inferred positions. We quantify this effect for cosmological estimates in section \ref{sec:cosmology}.

For an indication of which lines are important for different 21~cm observing redshifts, we decompose the power of each individual line in figure \ref{fig:n_space_contam}. For each 21~cm redshift, we compute the total contaminating RRL auto-power spectrum for each line indexed by quantum number $n$. This results in a distribution of contamination, bounded by those lines produced at $z = 6$ on the left (the maximum emitting redshift we consider in our calculation), and those at $z=0$ on the right. We neglect emission from higher redshifts because the star formation rate declines rapidly beyond $z\sim6$, making its contribution to the contamination small. At the right edge, there is a sharp uptick in each curve from the divergence of the $1/\alpha_{\perp}^2$ factor in eq. \ref{eqn:Pkdecomp}, driven by those lines that fall close to $z=0$. While the lines dominating the signal vary with 21~cm redshift, $130 < n < 250$ produce the strongest signal for all redshifts. The rest frame frequency of these lines corresponds to a 21~cm observation at the peak of star formation ($z_{21}\sim2$). For each of the curves, the 20 lines closest to $z_{\rm RRL}=0$ account for less than half of the total contamination to the 21~cm signal, ($\sum_{n(z_{\rm RRL = 0})}^{n+20}\delta\Delta^2_{\rm RRL}(n)<0.5$).

\begin{figure}
    \centering
    \includegraphics[width=1\linewidth]{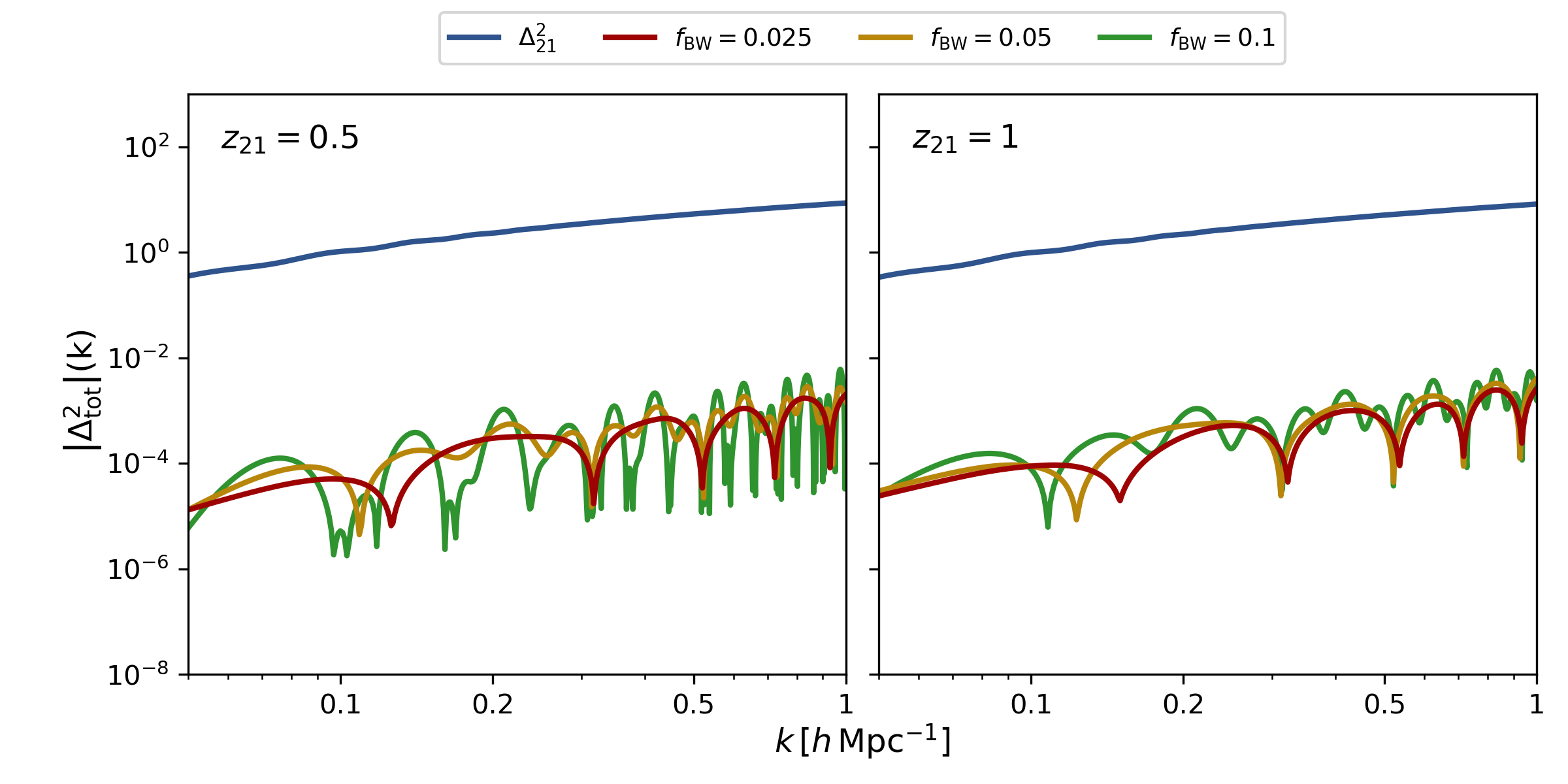}
    \caption{Total RRL power spectra for different observational bandwidths (parameterized as fractional redshift width, $f_{\rm BW}$) compared to the 21~cm auto-power spectrum (solid blue) at $z=0.5$ (left panel) and $z=1$ (right). At higher redshifts ($z\gtrsim2$), the spectra have minimal variation between bandwidths and no oscillatory features and so are not shown. All curves use Model C in table \ref{tab:RRLModels} and are calculated with $k_\perp = 0$.}
    \label{fig:PSPLOT_BW}
\end{figure}

The differences in Model C and D arise from the shape of their cumulative optical depth curves (see figure \ref{fig:tauCompare}). We find those models with a lower density (in this case, Model C) results in the strongest contamination arising from higher $n$ RRLs produced at lower redshifts, whereas higher density models are dominated by higher redshift (lower $n$) contamination. In our higher-density models, free-free absorption also suppresses a significant fraction of the highest-redshift emission.

Finally, in figure~\ref{fig:PSPLOT_BW} we examine the impact observational bandwidth has on the stacked RRL power spectrum. We choose bandwidths that correspond to the BAO scale ($k\sim 0.05-0.1~ h \rm~ Mpc^{-1}$) as this is the primary interest for this body of work \citep{Villaescusa_Navarro_2016}. At $z_{21}=0.5$ (left panel), for a narrow bandwidth (solid red curve) individual lines must be very close in frequency for the same structures to be seen in both lines (e.g. two RRL lines or a RRL and a 21~cm line). When lines that are closer in frequency dominate the cross terms, this leads to broader oscillations. As the bandwidth increases, individual RRL emitters that are farther apart in redshift space contribute to the RRL-RRL cross-power spectrum, which introduces more oscillations into the stacked power spectrum. At the widest bandwidth (green curve), the lines capable of contaminating the 21~cm auto-power spectrum can be much farther apart in redshift space, resulting in a higher oscillation frequency. As the redshift increases ($z_{21}=1$, right panel) and surface brightness dimming dominates, the oscillations are suppressed as the RRL auto-power spectrum begins to take over. For $z_{21} \gtrsim 2$, the shot noise term dominates $|\Delta^2_{\rm tot}|$, so the combined power spectrum smooths out and has limited oscillatory features, and so we have omitted those two panels from figure \ref{fig:PSPLOT_BW}.

\begin{figure}
    \centering
    \includegraphics[width=0.9\linewidth]{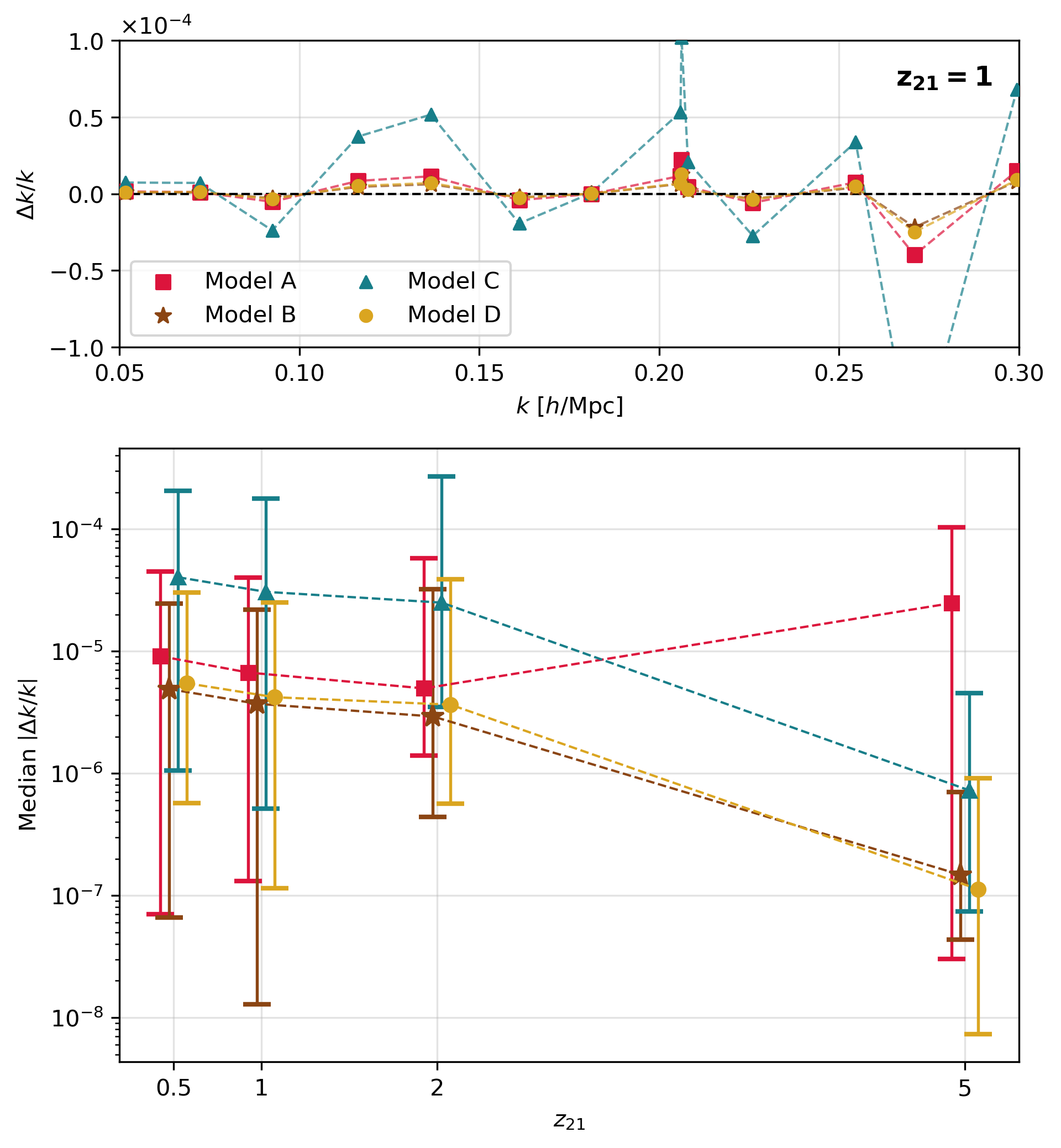}
    \caption{Shift in the extrema of the BAO due to the inclusion of RRLs in the 21~cm power spectrum with $k_{\perp}=0$. \textit{Top:} The shift for each RRL model in figure \ref{fig:tauCompare} for the 21~cm redshift of $z_{21}=1$. We bound the x-axis to focus only on those $k$ values of interest in modern 21~cm BAO experiments.
    \textit{Bottom:} Each point shows the median absolute fractional shift in the BAO extrema within the relevant range ($0.05 < k < 0.3~h\,{\rm Mpc}^{-1}$), while the bars span the maximum to minimum value for the corresponding redshift and model within that range. The bars are slightly offset from their redshift for visualization purposes. For all physical models and redshifts, the shift in the BAO extrema is below the level set by cosmic variance of $\sim10^{-3}$.}
    \label{fig:cosmology}
\end{figure}

Together, these results demonstrate that the RRL emission is structured, evolves with redshift, remains consistent across bandwidths and optical depth models, and may be large enough to bias future 21~cm intensity mapping observations and precise BAO estimates. We evaluate this bias in the next section and find it is small enough to not be a concern.
\subsection{Cosmological Implications}\label{sec:cosmology}
Forecasts for future 21~cm BAO estimates target subpercent precision, and so small fluctuations can significantly impact cosmological parameter estimates.
In particular, contamination that generates oscillatory features can shift the inferred location of the BAO extrema. Even part-in-a-thousand shifts could lead to biases in inferred parameters that are comparable to the statistical error. In this section, we calculate the shift in the BAO extrema in the 21~cm intensity mapping power spectrum from the contamination predicted by our four calibrated HRRL models.

We quantify this shift to the BAO extrema, including both peaks and troughs, with a perturbative approach using Newton's method, where we begin by fitting a third-order polynomial to both the `clean' 21~cm power spectrum, and to the stacked RRL power spectrum $P_{\rm contam} = P_{\rm RRL} + P_{\rm 21-RRL} + P_{\rm RRL-RRL}$. Subtracting these smoothed fits isolates the oscillatory BAO component of the power spectra. The position of the BAO extrema $k_0$ is defined as the location of the maximum of the uncontaminated 21~cm power spectrum,
\begin{equation}
    \frac{dP_{21}}{dk}\Bigr|_{k_0} = 0.
\end{equation}
When including RRL contamination, the observed power spectrum becomes $P_{\rm tot}=P_{21}+P_{\rm contam}$, and including a smaller perturbation $\Delta k$ gives a power spectrum with a shift
\begin{equation}
\frac{dP_{\rm tot}}{dk}\Big|_{k_0 + \Delta k} = 0.
\end{equation}
Expanding this to second order in $\Delta k$ and assuming that the contamination is small gives
\begin{equation}
    0 = \underbrace{\frac{dP_{21}}{dk}(k_0)}_{0} + \Delta k\frac{d^2P_{21}}{dk^2}(k_0) + \frac{dP_{\rm contam}}{dk}(k_0) \rightarrow \Delta k = -\frac{dP_{\rm contam}/dk}{d^2P_{21}/dk^2}
\end{equation}
Since the real quantity of interest is the fractional shift in this BAO scale ($\Delta k / k_0$), we plot that in figure \ref{fig:cosmology}.  
We perform this analysis on the fixed optical depth model used in \citetalias{Petrovic_2011} ($\tau_{\rm RRL} = 0.1$) and find the shift in the BAO extrema for low redshifts is $\Delta k / k_0 \approx 10^{-3}$ at $z_{21}=0.5$. However, the perturbative method breaks down at $z_{21}\gtrsim2$ (where $\Delta k / k_0 \gtrsim 1$) as the amplitude of the contamination exceeds the 21~cm signal (see figure~\ref{fig:PSPLOT_resid}) and so supports the interpretation of an extreme upper-bound comparison rather than a physically motivated prediction.

For our physically-motivated models in figure \ref{fig:tauCompare}, we find the introduction of $P_{\rm contam}$ produces a shift for all relevant wavenumber and at all redshifts. The upper panel of figure \ref{fig:cosmology} shows the fractional shift for each individual $k_0$ extremum at redshift $z_{21}=1$. The shifts vary in both sign and magnitude across the relevant range $0.05 < k_0 < 0.3~h\,{\rm Mpc}^{-1}$, but remain small for all four of our models. 

The bottom panel of figure \ref{fig:cosmology} summarizes this effect for all our models and across all redshifts. We find the median fractional shift in the BAO extrema of a few $\times 10^{-6}$ to a few $\times 10^{-5}$ at redshifts $z_{21}\lesssim2$ for all our optical depth models. For $z_{21}=5$, Models B--D predict the shift in the extrema to be smaller by roughly one order of magnitude as the contamination is dominated by the smoother RRL auto-power spectrum term (as the RRL emission from similar redshifts as 21~cm is smaller at such high redshifts in our models). Model A is the exception to this, likely resulting from the larger contribution from high $n$ RRLs.
Future BAO analyses aim for precision down to the limit set by cosmic variance at the $10^{-3}$ level, and we find the contamination in our models is well below the threshold where it poses a significant problem \citep{Oxholm_2021}. We note our fiducial model is observationally motivated (see figure \ref{fig:tauCompare}) and can likely be interpreted as an upper bound on what is possible.

\section{Conclusions}
\label{sec:Conclusions}

We have forecast the level of radio recombination line (RRL) contamination of post-reionization 21~cm intensity mapping surveys.  We developed a simple model for RRL emission, building off the work of \citetalias{Petrovic_2011} -- who investigated the contamination of the much brighter reionization-era 21~cm signal -- and parameterizing the emission by the frequency-integrated optical depth and covering fraction of the galaxy's radio continuum. We estimated the level of RRL contamination of the 21~cm power spectra, as well as the position of the BAO features.

Our main results are as follows:
\begin{itemize}

    \item We updated calculations for the hydrogen RRL brightness temperature that appeared previously in the literature with physically-motivated models, informed by the physics of galactic $\hii$ regions and calibrated to RRL observations (see figure \ref{fig:tauCompare}). We presented the fractional contamination to the 21~cm brightness temperature signal, finding that all of our physically motivated models (but not the \citetalias{Petrovic_2011} model) result in a percent-level or smaller contamination to the mean 21~cm brightness temperature at all redshifts (see figure~\ref{fig:contamSurv}). 

    \item We estimated the contribution of carbon RRLs to the cosmological 21~cm brightness temperature emitted by cold diffuse gas. For our fiducial CRRL parameters, the resulting contamination is smaller than that of our HRRL models and so we expect CRRLs to be subdominant in post-reionization 21~cm intensity mapping.

    \item We developed the formalism for calculating the RRL auto- and cross-power spectra, finding the contamination to the 21~cm auto-power spectrum is minimal at all redshifts. We find the contributions enter primarily from the cross-power terms resulting from spatially proximate emission that scale as $\overline T_b^{\rm RRL}/\overline T_b^{21}$, rather than the auto-power terms that scale as $\left(\overline T_b^{\rm RRL}/\overline T_b^{21}\right)^2$. 

    \item RRLs that are only \emph{slightly} offset in frequency from the 21~cm line (or each other) produce highly oscillatory features in the total combined RRL power spectrum owing to cross terms between these lines (see figures~\ref{fig:PSPLOT_indiv} and \ref{fig:PSPLOT_resid}). Our models predict contamination near BAO scales ($k \sim 0.1~ h \rm~ Mpc^{-1}$) on the order of $10^{-4}P_{21}$ at all redshifts. At $z_{21}\leq2$, the highly oscillatory cross-terms dominate the RRL contamination, while at $z_{21}=5$ the contamination is dominated by the smoother RRL auto-power spectra summed over the many contaminating lines.

    \item The dominant contribution to the HRRL auto-power spectrum at $k=0.1 ~h~\rm Mpc^{-1}$ comes from transitions with quantum number $130\lesssim n \lesssim 250$ (see figure~\ref{fig:n_space_contam}). We find that the effective density of extragalactic $\hii$ regions shapes which lines contribute, with lower-redshift (higher-$n$) emission dominating the signal for our models with $n_e=10^3$~cm$^{-3}$, whereas higher-redshift (lower-$n$) emission dominates for denser $\hii$ regions with $n_e=10^4$~cm$^{-3}$. In these higher-density models, free-free absorption also suppresses a significant fraction of the highest-redshift emission.

    \item We show that the observational bandwidth impacts the frequency of oscillations that appear in the power spectrum, due to the increased number of RRLs that can fall within a wider frequency band from the same structures as the 21~cm signal. The increased number of lines in an observing band imprints slightly higher frequencies in the acoustic-like oscillations in the contaminating cross power spectrum terms (see figure~\ref{fig:PSPLOT_BW}).

    \item We find the RRL emission biases the location of the BAO extrema that would be inferred by 21~cm intensity mapping experiments by a small amount at all redshifts, with $\Delta k/k$ at most a few $\times 10^{-4}$ (see figure~\ref{fig:cosmology}). This is lower than the limit set by cosmic variance by about an order of magnitude \citep{Oxholm_2021}. The choice of RRL model does not dramatically impact this result and so our fiducial Model C can be thought of as an upper bound.
\end{itemize}

We note our empirical calibration for the RRL optical depth models mostly use lower-$n$ RRL observations of low-redshift galaxies, 
with only two measurements reported at $z>0.5$. We also find that the dominant contamination comes from relatively high quantum numbers that emit near the 21~cm rest frame frequency ($n \sim 170$, see figure~\ref{fig:n_space_contam}).  High-redshift galaxies may be more compact, allowing larger covering fractions of $\hii$ regions, or have denser $\hii$ regions owing to higher ISM pressures, and both factors could increase their RRL emission relative to lower redshift galaxies.  However, since our models consider a wide range of $\hii$ densities, we expect that it would be challenging to devise a model that exceeds our predictions for the RRL contamination of $\sim 10^{-4}P_{21}$ by more than an order of magnitude. Moreover, oscillations that align more closely with the BAO peaks and troughs could produce a larger shift in the inferred BAO scale.

In this work we find that the level of RRL contamination is not significant even for the most ambitious upcoming 21~cm efforts. However, for scenarios with larger RRL contamination, cross-correlations with other large-scale structure tracers can be used to detect the effect of RRLs in 21~cm intensity mapping surveys (e.g., \cite{Cheng_2022, Fronenberg_2024}). Such correlations could be used to develop templates for the contamination and remove its effect to reduce biases in cosmological parameter measurements.

\section*{Acknowledgments}
We thank Adam Lidz for directing us towards the work by \citetalias{Petrovic_2011}, which motivated much of this work. We thank Simon Foreman, Adam Lidz, and Jordan Mirocha for their valuable feedback on the manuscript. We also thank Kimberly Emig and Pedro Salas for their helpful discussion and contribution of departure coefficient tables and code used in our analysis. This material is based upon work supported by the National Science Foundation Graduate Research Fellowship Program under grant number DGE-2140004. PP would like to acknowledge the ARCS Foundation for their generous support. PP would like to thank his Qualifying Exam Committee of Eric Agol and Jim Davenport for their comments and valuable feedback on the paper draft. We gratefully acknowledge the Pacific Postdoctoral Program at the Dark Universe Science Center, University of Washington, during which part of this work was carried out. The Pacific Postdoctoral Program is supported by the Simons Foundation (SFI-MPS-T-Institutes-00012000, ML).

\bibliographystyle{aasjournal}
\bibliography{bib}
\end{document}